\documentclass[10pt,journal,compsoc]{IEEEtran}

\usepackage[utf8]{inputenc} 
\usepackage[T1]{fontenc}    
\usepackage{hyperref}       
\usepackage{url}            
\usepackage{booktabs}       
\usepackage{amsfonts}       
\usepackage{nicefrac}       
\usepackage{microtype}      
\usepackage{lipsum}
\usepackage{fancyhdr}       
\usepackage{graphicx}       
\graphicspath{{media/}}     
\usepackage{xcolor}
\usepackage{todonotes}
\usepackage{amsmath}
\usepackage{enumitem}

\usepackage{subfig}

\usepackage{pifont}
\usepackage{fontawesome}
\usepackage{pdflscape}
\usepackage{afterpage}
\usepackage{threeparttable}
\usepackage{supertabular}

\usepackage[acronym, shortcuts, nohypertypes={acronym}]{glossaries}
\newacronym{AC}{AC}{audio captioning}
\newacronym{AI}{AI}{artificial intelligence}
\newacronym{ALM}{ALM}{audio-language model}
\newacronym{ASC}{ASC}{acoustic scene classification}
\newacronym{ASR}{ASR}{automatic speech recognition}
\newacronym{AST}{AST}{audio spectrogram transformer}
\newacronym{BERT}{BERT}{bidirectional encoder representations from transformers}
\newacronym{CALM}{CALM}{composition to augment language models}
\newacronym{CNN}{CNN}{convolutional neural network}
\newacronym{DCASE}{DCASE}{detection and classification of acoustic scenes and events}
\newacronym{DL}{DL}{deep learning}
\newacronym{DNN}{DNN}{deep neural networks}
\newacronym{EMA}{EMA}{exponential moving average}
\newacronym{EMR}{EMR}{elect, mask \& rescale}
\newacronym{EOS}{<EOS>}{end-of-sequence}
\newacronym{FM}{FM}{foundation model}
\newacronym{GMM}{GMM}{Gaussian mixture model}
\newacronym{LLM}{LLM}{large language model}
\newacronym{LMC}{LMC}{linear mode connectivity}
\newacronym{LoRA}{LoRA}{low-rank adaptation}
\newacronym{LSTM}{LSTM}{long short-term memory}
\newacronym{MAVD}{MAVD}{Montevideo Audio and Video
Dataset}
\newacronym{ML}{ML}{machine learning}
\newacronym{MLP}{MLP}{multi-layered perceptron}
\newacronym{NLP}{NLP}{natural language processing}
\newacronym{PANN}{PANN}{pretrained audio neural network}
\newacronym{PaSST}{PaSST}{patchout fast spectrogram transformer}
\newacronym{PEFT}{PEFT}{parameter efficient fine tuning}
\newacronym{RAG}{RAG}{retrieval-augmented generation}
\newacronym{RIR}{RIR}{room impulse response}
\newacronym{SED}{SED}{sound event detection}
\newacronym{SER}{SER}{speech emotion recognition}
\newacronym{SOS}{<SOS>}{start-of-sequence}
\newacronym{SSL}{SSL}{self-supervised learning}
\newacronym{STFT}{STFT}{short-time Fourier transform}
\newacronym{SVM}{SVM}{support vector machine}
\newacronym{VLM}{VLM}{vision-language model}
\newacronym{ViT}{ViT}{vision transformer}

\usepackage{soul}
\newcommand{\add}[1]{#1}
\newcommand{\remove}[1]{}
\newcommand{\change}[2]{#2}

\newcommand{\wav}{\emph{wav2vec2.0}}
\newcommand{\cf}{{c.f.,\ }}

\usepackage[natbib, bibencoding=utf8, citestyle=numeric, bibstyle=ieee, maxbibnames=999, maxcitenames=2, mincitenames=1, sortcites]{biblatex}
\bibliography{references}

\usepackage[capitalize]{cleveref}

\begin{document}
\title{Computer Audition: From Task-Specific \\Machine Learning to Foundation Models}

\author{
  Andreas~Triantafyllopoulos,
  Iosif~Tsangko,
  Alexander~Gebhard,\\
  Annamaria~Mesaros,
  Tuomas~Virtanen,
  Björn~W.~Schuller
    \IEEEcompsocitemizethanks{\IEEEcompsocthanksitem A. Triantafyllopoulos, I. Tsangko, A. Gebhard, and B.\,W. Schuller were with the Chair of Health Informatics, Technical University of Munich (CHI), MRI, Munich, Germany, the Chair of Embedded Intelligence for Health Care and Wellbeing (EIHW), University of Augsburg, Augsburg, Germany, and the Munich Center for Machine Learning (MCML), Munich, Germany.
    \IEEEcompsocthanksitem A. Messaros and T. Virtanen were with the Audio Research Group, Tampere University, Tampere, Finland.
    \IEEEcompsocthanksitem B.\,W. Schuller was additionally with the Group on Language, Audio (GLAM), \& Music, Imperial College, London, UK and the Munich Data Science Institute (MDSI), Munich, Germany.
    \IEEEcompsocthanksitem Correspondence: andreas.triantafyllopoulos@tum.de
    }
    \thanks{Manuscript received XXX; revised XXX}
}

\markboth{Accepted for publication in PIEEE}%
{Triantafyllopoulos \MakeLowercase{\textit{et al.}}: Computer Audition from Traditional to Foundation Models}

\IEEEtitleabstractindextext{%
\begin{abstract}
    Foundation models (FMs) are increasingly spearheading recent advances on a variety of tasks that fall under the purview of computer audition -- \add{i.\,e.,} the use of machines to understand sounds.
    They feature several advantages over traditional pipelines: among others, the ability to consolidate multiple tasks in a single model, the option to leverage knowledge from other modalities, and the readily-available interaction with human users.
    Naturally, these promises have created substantial excitement in the audio community, and have led to a wave of early attempts to build new, general-purpose foundation models for audio.
    In the present contribution, we give an overview of computational audio analysis as it transitions from traditional pipelines towards auditory foundation models.
    Our work highlights the key operating principles that underpin those models\remove{,} and showcases
    how they can accommodate multiple tasks that the audio community previously tackled separately.
\end{abstract}
\begin{IEEEkeywords}
Computer Audition, Computational Audio Analysis, Foundation Models, Large Audio Models, Machine Listening, Acoustic Scene Classification, Sound Event Detection, Audio Captioning, Artificial Intelligence
\end{IEEEkeywords}}

\maketitle

\IEEEraisesectionheading{\section{Introduction}\label{sec:introduction}}

\IEEEPARstart{S}{ound} is everywhere around us.
A bird singing to its mate, the hiss of an espresso machine, the buzz of an insect in the afternoon sun, the drip of rain on a metal rooftop, or, belatedly, the angry horn of a passing driver, are some examples of the soundscapes that humans encounter in their daily lives.
These stimuli carry an enormous amount of information, from the state that their sources are in, to their location, and to their importance to the happenstance listener.
We process that information instinctively; while some of the other senses can be blunted at will, sound is omnipresent, a vital source of information and constant joy, or nuisance, to our ears.

It stands to reason that machines must be able to listen as well.
This ability can unlock a number of critical \emph{affordances} for artificial \emph{agents}\add{, i.\,e., artificial entities that interact with the physical or digital world,}\remove{ that operate in the real world}\remove{ --} especially if those agents are embodied.
Accordingly, \emph{computer audition} can provide actionable information regarding the surrounding environment \add{which can be used by the agent itself or a human overseer when planning future actions}.
It can also help an agent orient itself and keep track of entities in its environment even after they have moved beyond its direct field of vision.
These capabilities can be extremely useful for a wide range of applications, including Internet of Things devices that passively monitor their environment, autonomous driving agents that require a very dense, and temporally accurate, awareness, or even intelligent assistants who benefit from additional context.

Imparting the ability to listen onto machines has long been the purview of computational audio analysis research~\citep{Brown94-CAS} -- the field of study concerned with how machines can listen as good, or better, than humans.
However, for many decades, research efforts have been splintered among its different subfields, with custom, ad-hoc solutions to increasingly niche problems\add{, ranging from classifying general soundscapes~\mbox{\citep{Mesaros16-TUT}} to recognising sounds in very restricted domains, such as monitoring machines for malfunctions~\mbox{\citep{Dohi22-MDS}} or tracking traffic~\mbox{\citep{Damiano24-CSD, Zinemanas19-MAVD}}}.
That standard mode of operation is quickly changing, swept along by the tide that is transforming the wider \ac{AI} community.
In recent years, we are seeing an increasing trend towards \emph{consolidation}, with \emph{\ac{DL}} first leading the charge towards a common set of neural architectures capable of performing well on an array of different tasks, and, more recently, \emph{\acp{FM}} taking this trend to its extreme \change{--}{by pursuing} a single, `omnipotent' model capable of tackling multiple tasks~\citep{Bommasani21-OTO}.

Recent surveys have documented this trend~\citep{Latif23-SOL, Wan24-KNO, Caffagni24-REV, Yin23-SUR, Liu22-ASL, Awais25-FMD}.
However, they adopt a primarily `descriptive' approach, focused on diligently documenting new architectures, datasets, models, training regiments, and other information that helps characterise and differentiate between different \acp{FM}.
We opt for a different perspective.
\textbf{Our contribution aims to elucidate the inner workings of \acp{FM} for audio and ground them in the specific tasks relevant for computer audition}.
We thus aim to bring together two communities: the broader \ac{AI} public working on improving \acp{FM} without a clear emphasis on computer audition, and audio researchers who have insofar focused on specific subproblems of the field.
Accordingly, we begin our overview with a presentation of common tasks and conventional training methods.
This serves to ground our conversation and provide a comparison point for contemporary \acp{FM}.

We also note that, in this paper, we exclusively focus on `general' audio -- i.\,e., \emph{soundscapes}, \emph{audio scenes}, and \emph{sound events} -- that are distinct from speech or music.
Moreover, we use the terms \emph{computer audition} and \emph{computational audio analysis}\add{,} instead of computational auditory scene analysis (CASA~\citep{Brown94-CAS,Wang06-CAS}), to highlight the development of the field which has allowed addressing much more high-level processing tasks (e.\,g., recognising sources and describing them using natural language) than the CASA methods of the previous generation, which were aiming for more low-level tasks, such as F0 estimation, source separation, and perceptual modelling.

The remainder of our contribution is structured as follows.
We first describe the most common audio tasks in \cref{sec:tasks} and summarise traditional `best-practices' for dealing with each task in \cref{sec:conventional}.
Following that, we provide an overview of how audio-language \acp{FM} work in \cref{sec:afm}.
\cref{sec:frontiers} outlines ongoing research efforts and sketches out the next frontiers for this line of work, while \cref{sec:conclusion} contains a short summary and conclusion.

\section{Computer audition}
\label{sec:tasks}
    
    

\begin{table*}[t]
    \caption{
    Overview of the different computer audition tasks we are considering here.
    }
    \label{tab:applications}
    \centering
    \begin{tabular}{p{.4\textwidth}|p{.6\textwidth}}
    \toprule
       \textbf{Task} & \textbf{Description} \\
       \midrule
       \midrule
        \textbf{Acoustic scene classification} & Corresponds to the classification of an entire soundscape, treating all the sound sources in a scene as a whole. A single-label, multi-class, clip-level classification problem. \\
        \midrule
        \textbf{Sound event classification} & Corresponds to the classification of a single sound source, assuming that only that source is present in an audio clip. A single-label, multi-class, clip-level classification problem.\\
        \midrule
        \textbf{Audio tagging} & Corresponds to the classification of multiple, concurrent sound sources, assuming that more than one sources are present in an audio clip. A multi-label, clip-level classification problem.\\
       \midrule
        \textbf{Sound event detection} & Corresponds to the identification of multiple sound sources, as well as their temporal activities. A multi-label, segment-level classification problem.\\
       \midrule
        \textbf{Sound state and trait prediction} & Corresponds to the identification of a particular state or trait which characterises a source in a given moment, as well as (optionally) the actual source in question. A multi-label classification problem. Can be either clip- or segment-level.\\
       \midrule
        \textbf{Audio captioning} & Corresponds to the generation of a complete sentence characterising an audio soundscape and some or all of its sources. A sequence-to-sequence, clip-level classification problem. \\
       \midrule
        \textbf{Audio question answering} & Corresponds to either the generation of a complete sentence or the selection (i.\,e., classification) of the correct answer from a set of candidate answer in response to a query relating to an audio file. Either a classification or a sequence-to-sequence problem. Can be clip-level or segment-level. \\
       \midrule
        \textbf{Spatial localisation and tracking} & Corresponds to the estimation of a source's location in physical space, and its change in time. Optionally includes the identification of the type of source. Typically a segment-level regression problem. \\
       \midrule
        \textbf{Environmental parameter estimation} & Corresponds to the estimation of physical parameters of space (e.\,g., room size). Typically a clip-level regression problem. \\
    \bottomrule
    \end{tabular}
\end{table*}

An overview of the basic audio tasks is presented in \cref{tab:applications}.
We describe each one in more detail below, focusing on its motivation and importance.
Additionally, given the data-driven nature of audio research today, we highlight the most commonly-used datasets for each task and the challenges that arise from those, which adds some nuance to the capabilities and promises of models that use them.
Finally, we highlight the \emph{challenges} that traditional methods face and the \emph{opportunities} of \acp{FM} to disrupt them.

\subsection{Acoustic scenes and events}

\textbf{Task description:} The most basic tasks for everyday audio analysis are \ac{ASC}\add{~\mbox{\citep{Barchiesi15-ASC}}}, aiming to classify the whole scene into one class, and \ac{SED}\add{~\mbox{\citep{Mesaros21-SED}}}, aiming to detect individual sound events and, optionally, their temporal activity.
We note the difference between the more challenging task of \ac{SED} and the more limited scope of audio tagging -- the multi-label prediction of sound presence \emph{anywhere} within a given audio clip, i.\,e., without caring for temporal information -- and sound even\add{t} classification -- the multi-class prediction of a single sound, again without attention to time.
All the above tasks are motivated by the fact that sounds provide information about events happening in the overarching environment.
Therefore, an \change{intelligent}{artificial} agent that has access to that information can appropriately react to those events.
The development of automatic classification methods for everyday audio is thus largely driven by the need of identifying such contexts and sources of sound, with the main envisioned applications being in-context awareness for devices and autonomous agents. 
This follows human evolution, where the passive listening of the environment in which categorisation of the sounds is the main purpose of the perception process~\citep{Lyon17-HAM}.

\textbf{Datasets:} With this goal in mind, the datasets available for training such classification systems are aimed at characterising soundscapes in terms of sound sources -- who or what is making the sound -- and, sometimes, the action -- how the sound is produced -- resulting in labels like ``bird singing'', ``people talking'', or ``car passing by''~\citep{Barchiesi15-ASC}.

There are many datasets available for those tasks, but most of them are focused on a few event classes and feature data from a restricted set of recording environments.
The largest one for audio tagging is the AudioSet dataset published by Google~\citep{Gemmeke17-ASA}, containing over 500\remove{,} partially overlapping sound event classes, with a hierarchical ontology loosely based on WordNet\add{\footnote{\url{https://wordnet.princeton.edu/}}}.
The dataset consists of 10\,s segments extracted from video clips on YouTube, with the labels provided at clip level (weak labels).
A strongly-labelled subset (AudioSet strong)~\citep{Hershey21-BEN} was published later, containing temporal information for the sound events and can be used for \ac{SED}.
Due to the different annotation procedures, AudioSet strong has a rather different distribution/density of labels per clip, and generally better-quality labels.
Other datasets for \ac{SED} are TUT Sound Events~\citep{Mesaros16-TUT}, consisting of manually annotated recordings of 3-5\,min length and \change{DESED}{the Domestic environment sound event detection (DESED) dataset}~\citep{Turpault19-SED}, manually annotated clips of 10\,s length -- though, in comparison to AudioSet, both contain much less data-points.
Beyond AudioSet, datasets for audio tagging or sound event classification are typically based on Freesound, such as FSD50k~\citep{Fonseca21-FAO}, FSD18knoisy~\citep{Fonseca19-LSE}, and ESC-50~\citep{Piczak15-ESC}. 
For \ac{ASC}, the largest datasets are the TAU Urban acoustic scenes 2019 dataset (40\,h of data)~\citep{Mesaros18-AMD}, with an audio-visual subset of 34\,h~\citep{Wang21-ACD}, containing recordings of public locations from 12 large European cities, and the Cochlscene dataset~\citep{Jeong22-CAO}.
TAU UASC has been provided and is still used in the \ac{DCASE} challenge\add{\footnote{\url{https://dcase.community/}}}, with an evaluation set still not publicly available\add{\footnote{See \mbox{\citet{Mesaros18-ASC}} for an overview of challenge results for 2017.}}.



\textbf{Challenges:} While the performance on classification tasks keeps improving year after year, there are a few significant issues that remain challenging.
First of all, the most popular setting of a classification task, be it for \ac{ASC} or \ac{SED}, aims to predict a flat ontology, in which sounds either belong or do not belong to a particular class.
This is common even though AudioSet, the most popular dataset to (pre)train on, is a hierarchical collection.
A few attempts at hierarchical classification do exist~\citep{Xu16-HLF, Jati19-HLF}, but they are not widely used or pursued in literature.

Another significant factor related to ontologies is the definition of the categories themselves.
Most often, sounds are annotated based on the source that produces them, which can result in a collection of sounds that are not necessarily similar from an acoustic point of view.
An ontology based on labels will create a hierarchy where high acoustic similarity can exist between classes that are not in any way semantically related, or create children of a parent class that are acoustically very different from each other (e.\,g., human sounds containing subclasses like speech, cough, breathing, sneezing, etc.).
For example, as mentioned, the AudioSet~\citep{Gemmeke17-ASA} labels are derived from WordNet, which emphasises semantic similarity, rather than acoustic similarity.
These factors create significant limitations in what can be produced in terms of application, with very high performance possible for well-defined tasks that target a reduced number of sufficiently different classes, but which are far from a general sound classification system that could deal with ambiguous sounds. 

\textbf{Opportunities:} The most recent developments in natural language processing are finding their way into neighbouring fields, including audio, bringing the possibility of explaining the content of audio in free-form natural language rather than just event class or acoustic scene labels.
Essentially, this might mean the abandonment of this traditional, coarse classification of sound in favour of the more nuanced approaches outlined below.

\subsection{From events and classes to states and traits}
\label{ssec:traits}

\textbf{Task description:} As discussed above, the field of audio analysis has been dominated by the identification of \emph{sound sources} and their organisation in \emph{scenes}.
This has been largely driven by the impetus to first decompose a soundscape into its core constituents before proceeding with a more granular analysis.

\begin{figure}[t]
    \centering
    \includegraphics[width=\columnwidth]{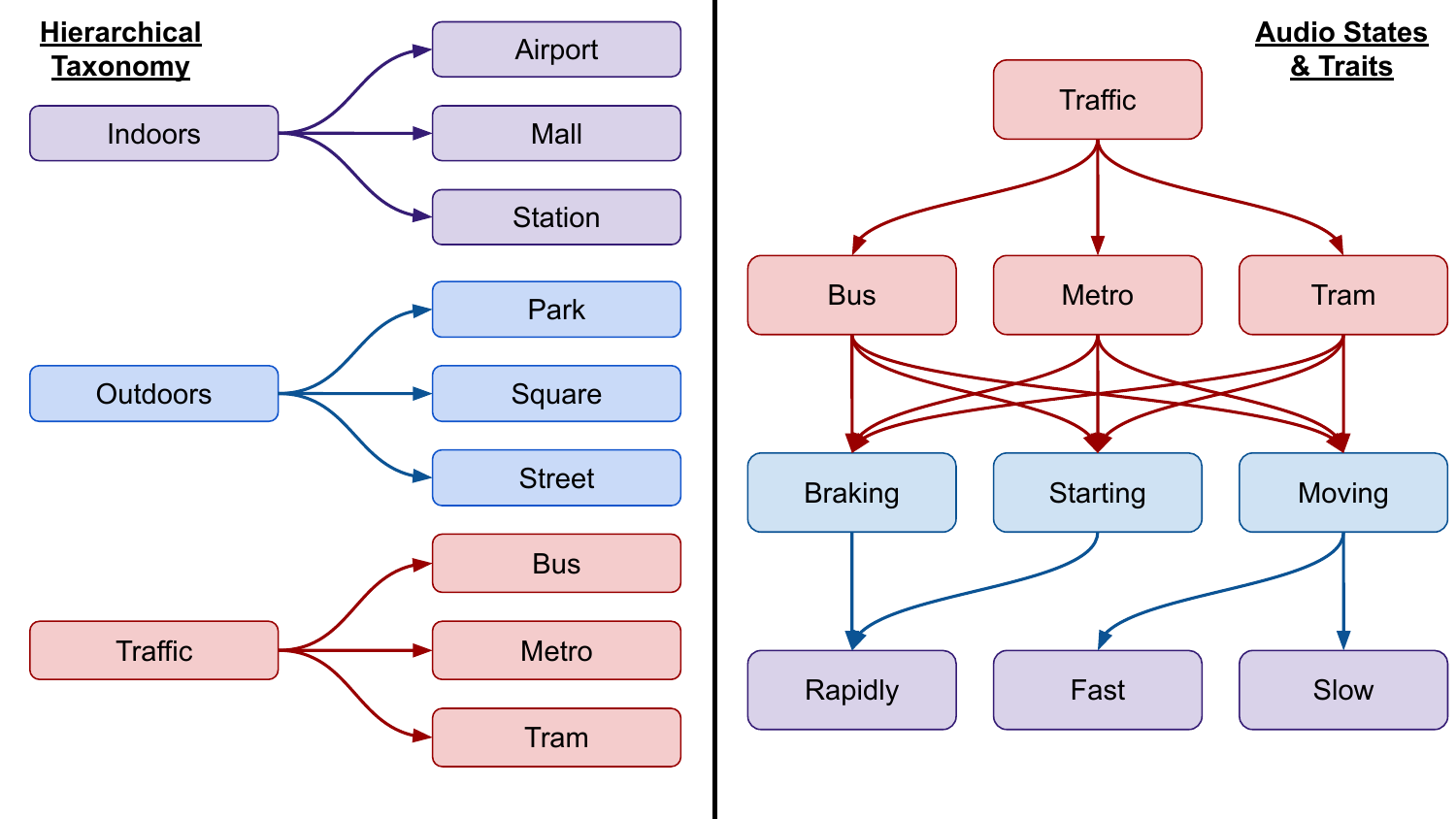}
   \caption{
   Contrasting a hierarchical organisation of sounds (left; inspired by TAU UASC) to a taxonomy that adopts states and traits (right; inspired by MAVD).
   Whereas a hierarchy follows a strict top-down approach, where every sound can trace a unique genealogy back to a root node, and no two different root nodes share common leaves, in states and traits we have a flexible set of leaf nodes that can be used to describe multiple nodes.
   For instance, in the case of traffic sounds, we have multiple vehicles (roots) sharing the same state (starting, accelerating, braking), and each of those states additionally sharing a number of attributes (fast, slow).
   In contrast, a strict hierarchy may classify sounds according to their origin (e.\,g., coming from an indoor or outdoor location) with no overlap between the two high-level categories.
   }
   \label{fig:traits}
\end{figure}

However, this operationalisation ignores that physical objects may appear as different manifestations (\emph{traits}) or be in different \emph{states} during different times.
For example, a car generates sound when starting, accelerating, coasting, and braking -- or crashing.
These constitute different \emph{states} of the same underlying class.
Moreover, a car may be new or old, \change{Diesel}{diesel} or electric, a sports car or a family station wagon, or have any number of other immutable or slow-changing \emph{traits} which alter the way it sounds.
Interestingly, these two different characterisations modulate one another: a new car accelerates, coasts, and brakes differently than an older car (even for the exact same model).
Therefore, states and traits can be combined compositionally into a spectrum of different sub-categories that all stem from a single object.
\cref{fig:traits} contrasts the two operationalisations.

Contemporary research often ignores these nuances when categorising sounds~\citep{Schuller21-NAI}.
The motivation is largely pragmatic: annotating a dataset for all potential states and traits for all objects is prohibitively expensive.
As a result, this is usually done for very constrained environments, such as detecting malfunctioning machines~\citep{Kawaguchi21-DAD} or monitoring traffic~\citep{Zinemanas19-MAVD} -- even there, the ontology is either very restricted or the problem is cast as one of anomaly detection.

States and traits, however, have important repercussions for computer audition systems.
Their purpose serves the creation of a more comprehensive \emph{ontology of sound}.
By expanding the aspects of an auditory scene that can be recognised, states and traits facilitate a more thorough understanding of the surrounding environment.
For example, knowing the difference between an accelerating and a braking vehicle can make a world of difference in an autonomous driving setup, especially since the auditory cue may be available faster than an estimation based on vision or depth estimates;
a malfunctioning motor might be a reason for concern for a robot placed on a factory floor;
a door knock that sounds urgent might prompt a more immediate response in an intelligent house assistant, 
and so on.
Understanding is a necessary prerequisite for higher-order reasoning, which is becoming more important given the expected pervasiveness of autonomous agents in daily life, and the role that audition has to play as part of their sensing systems.

States and traits can be further traced down to two different roots: a mechanical and a perceptual one.
The mechanical aspect of states and traits is dictated by the physical properties of the object in question.
Sound emerges from objects which exist in, or are temporarily brought to, a state of vibration.
During that process of vibration, sound waves are generated (and are then propagated through a medium, typically air or water).
The state of the sound generating object, as determined by its geometry and material property, will influence how those waves are generated and how they propagate through space within its boundaries, and, correspondingly, outside of them -- a half-empty bottle of water will sound more \emph{hollow} when thrown against the wall than a full one.
This constitutes an `objective ground truth' for how an object sounds in different contexts.

On the other end of those waves are human receivers
\footnote{We note that our perspective here is deliberately anthropomorphic.
Animals too can perceive sounds and may have different reactions to them.
However, for the purposes of this work, we stick to man ``as the measure of all things''.}.
The description and categorisation of sounds depends on the perceptual qualities that humans attribute to them.
From a physiological perspective, this is restricted by the sounds we can process and differentiate between.
For example, the human hearing system can only capture frequencies up to \add{ca.} \change{22,000}{20,000}\,Hz; it has a more granular perception of pitch in lower frequencies, etc.
Furthermore, there are cultural aspects at play when it comes to a human perception of sounds~\citep{Scott15-MHF, Dick18-OTI, Lemaitre18-AAP},
Prior research has shown that human listeners distinguish sounds partially based on familiarity, and have specific affective responses to them.
These are largely driven by prior experience and environment, and consequently are heavily dependent on culture and origin.

Finally, a particular aspect of sound generation that is important for characterising states and traits is \emph{intentionality}.
Sounds that are emitted by human-manufactured objects are oftentimes intentionally designed to elicit specific responses.
An alarm clock can be be perceived as more or less annoying depending on whether it is set to produce an urgent, high pitch ringing or a soft buzz (also, in both cases, on who needs to wake up and at what time).

\textbf{Datasets:} 
There are very few datasets annotated for the states and/or traits of the sound sources they contain, and most are limited to a very narrow domain.
One example is the \ac{MAVD}~\citep{Zinemanas19-MAVD}, which contains audio(visual) annotations of traffic events, which do not only correspond to the vehicle that produces them (e.\,g., ``car'' or ``bus'') but also to the state it is in (e.\,g., ``accelerating'' or ``braking'').
One other commonly-used dataset is the Real World Computing Partnership-Sound Scene Database (RWCP)~\citep{Nakamura00-ASD}, which contains sound events collected with different materials.
However, RWCP is small in size and contains sound events collected in laboratory conditions, whereas MAVD is limited to a very specific context.
Epic-sounds~\citep{Huh23-EAL} is perhaps the largest dataset which features a state/trait ontology, though again with a limited scope (human actions in the process of cooking).

\textbf{Challenges:} 
The high degree of subjectivity, and the large span of potential leaf nodes, make for substantial challenges in creating comprehensive ontologies of sounds and annotating large datasets according to their specifications.
This remains an open problem as the amount of attributes that may characterise a particular audio source is very large and context-dependent.
There is also a difference between `physiological' attributes which can be measured objectively (e.\,g., the type of material involved) and perceptual attributes which are subjectively evaluated by annotators (e.\,g., whether an action happened fast or slowly).
For these reasons, the task of predicting states and traits of audio events is still in its nascent stages.

\textbf{Opportunities:} The holistic characterisation of audio in terms of states and traits is a research area primed for innovation using multimodal foundation models.
The main reason is that states and traits are compositional: a particular object may have several of them (also concurrently) and different objects share subsets of the catalogue of states and traits that are relevant for sound.
This compositionality is hard to capture with traditional \ac{ML}, which relies on exhaustive datasets covering all possible combinations and with sufficient examples to facilitate generalisation.
Rather, this problem can be more efficiently tackled by leveraging vast amounts of general world knowledge, a small set of labelled examples covering a subset of all states and traits, and advanced reasoning capabilities to transfer that knowledge to previously unseen combinations in zero-shot fashion -- a feat which foundation models have shown to be capable of~\citep{Bommasani21-OTO}.

\subsection{Audio captioning and question answering}

\textbf{Task description:} Automatic audio captioning refers to methods that describe the content of an audio signal by providing \emph{a textual description} (caption) to characterise it.
Captions can generally be free-form text, facilitating the annotation of richer information about acoustic scenes, going beyond the fixed ontologies typically used in machine perception.
In addition, captions can describe relationships between entities, for example ``A dog wearing a chain collar runs towards a pool, swims through it, and returns, then shakes the water off.'' (an example caption from the Clotho dataset~\citep{Drossos20-CAA}).

However, natural signals often contain multiple sound sources and a rich variety of information.
In audio captioning, this raises the challenge of which content\remove{s} captions should focus on.
While there are potentially many ways to control or guide the caption generation process, an alternative solution is audio question answering \cite{Fayek20-TRV}, where a question presented using natural language is used as an additional input together with the audio, and a question-answering system is expected to output an answer.
Depending on the question type, the answer can be limited to a set of specific classes, or be free-form text.
In the simplest case, the answer is binary (yes / no).
Finally, the audio-question task can be inverted to provide linguistic queries that are answered by providing the associated audio clips, or segments, that correspond to it -- a task known as audio-language retrieval~\citep{Xie22-LAR}.

\textbf{Datasets:} While the use of free-form text allows expressing potentially any information in audio, in practice methods are limited by training datasets.
The vocabulary size in commonly used audio captioning datasets is 4369 in Clotho \citep{Drossos20-CAA} and 4724 in AudioCaps \citep{Kim19-AGC}.
Commonly used audio captioning datasets have been typically produced by crowdsourcing captions for existing environmental audio datasets; Clotho is based on Freesound, and AudioCaps is based on AudioSet.
For this reason, recent work has focused on automatically captioned data using \acp{LLM}~\citep{Mei24-WAC}.
Existing audio question answering datasets and studies are also limited to data consisting of generated acoustic inputs and question/answer pairs~\citep{Fayek20-TRV}, feature crowdsourced small-scale datasets~\citep{Lipping22-CAC}, or rely on machine pseudo-captions (using \acp{LLM}) to increase the size of annotated data~\citep{Gong24-LTA}.

\textbf{Challenges:} One of the major challenges in audio captioning is the multitude of different ways one can describe the contents of audio signals. Furthermore, what makes a good caption depends on the application and intended users of the generated captions. While in \ac{SED} or \ac{ASC} we can consider that there is a correct ground truth output or a set of reference annotations which are used for evaluating and training systems, for audio captioning there can be multiple correct outputs. This should be taken into account at least in the evaluation of captioning methods, but more research will most likely be needed to model application-specific needs of captioning systems.

\textbf{Opportunities:} Captioning and question answering are, perhaps, the tasks with the greater upside for foundation models, given their direct dependence on language capabilities.
They are both amenable to the use of generative language models, rather than discriminative models which aim to predict words from a restricted vocabulary or answer questions in a restricted domain (e.\,g., yes/no).
Moreover, captioning systems will benefit greatly from the increased fluency of \acp{LLM}, which can improve the appropriateness of the response independently of fidelity.



\subsection{Spatial information from audio}
\label{sec:spatial}

\textbf{Task description:} When a recording is made with multiple microphones, it is also possible to analyse the spatial properties of the scene based on inter-channel information.
The most common spatial audio task is source localisation~\citep{Grumiaux22-ASO}, which is often extended to time-varying estimation of locations, i.\,e., \emph{tracking}~\citep{Evers20-TLC}.
Recently, localisation has also been combined with the recognition of sound sources~\citep{Adavanne18-SOU}.
Beyond that, it is also possible to estimate other properties of an environment, such as the room size or reverberation time~\citep{Srivastava21-BLI}.
It is also noteworthy that with deep learning, it seems possible to estimate certain kinds of spatial information, such as the distance of sound sources, even from one-channel signals~\citep{Neri24-SPE}.

\textbf{Datasets:} Manual annotation of locations of sound sources is laborious, and, therefore, the existing training (and evaluation) datasets for sound event localisation are limited in the size and sound classes included, or have resorted to synthetic material.
Furthermore, existing datasets are limited to sounds captured in indoor environments, as those allow for capturing the audio or simulating the environment more easily.

For instance, the dataset of \citet{Shimada24-STA} consists of 7\,h of audio-visual material recorded in office-like environments with a spherical microphone array, where the reference locations of 13 sound event classes are obtained by a visual tracking system.
Alternatively, the simulated datasets of \citet{Politis20-ADA} and \citet{Nagatomo22-WEA} use measured impulse responses between sound sources and microphones, which is used to convolve isolated sound events to create their spatial images, and can be used to generate large quantities of audio data.
\add{Finally, spatial data can be synthesised using simulation toolkits, such as \emph{pyroomacoustics}{~\citep{Scheibler18-PAP}} or \emph{gpuRIR}{~\citep{Diaz21-GAP}}. These allow for the placement of arbitrary sources in (shoebox) rooms of different sizes, or even custom 3D-meshes, and with different material properties for reflective surfaces.}

\textbf{Challenges:} Current \ac{DL}-based approaches for spatial audio tasks are based on supervised learning using task-specific datasets, without any pretraining or transfer learning from other tasks.
A major limitation in these approaches is their specificity to the microphone array: methods trained with a specific array geometry do not produce meaningful results with other geometries, and therefore, it is significantly more difficult to transfer knowledge learnt from one dataset to another domain, or compile a single large-scale dataset which could be used to learn large-scale models that would generalise a wide range of domains.
Furthermore, general-purpose pre-trained representations or \ac{SSL} models that could be used for localisation related tasks are not yet widely utilised (even though several studies have used unlabelled spatial audio and visual data together for representation learning~\citep{Mo22-ACL, Morgado20-LRF}).
\add{Finally, while simulation toolkits provide a way to circumvent the scarcity of spatial recordings, their \emph{simulation-to-real} gap hampers generalisation to real-life recordings{~\citep{Grumiaux22-ASO}.}}

\textbf{Opportunities:} On the one hand, foundation models can improve the current state-of-the-art in spatial audio processing by co-opting the knowledge they have assimilated during (pre)training.
On top of that, they stand out from traditional approaches in their ability to accommodate additional sources of information (e.\,g., data about other sensors or information about the underlying scene) which can be extremely beneficial in the case of spatial audio.
For instance, accelerometer data obtained by movement sensors embedded in headphones can facilitate better temporal tracking and localisation of sound sources~\citep{Veiback20-SSL}; this data can be more easily incorporated in a \ac{FM} than in traditional algorithms, which need to be customised to accommodate them.
\add{The ways in which this information can be introduced to {\acp{FM}} is discussed below.}


\section{Conventional training methods}
\label{sec:conventional}

\subsection{Supervised classification}

\begin{figure*}[t]
    \centering
    \subfloat[Acoustic scene classification]{\includegraphics[width=.48\textwidth]{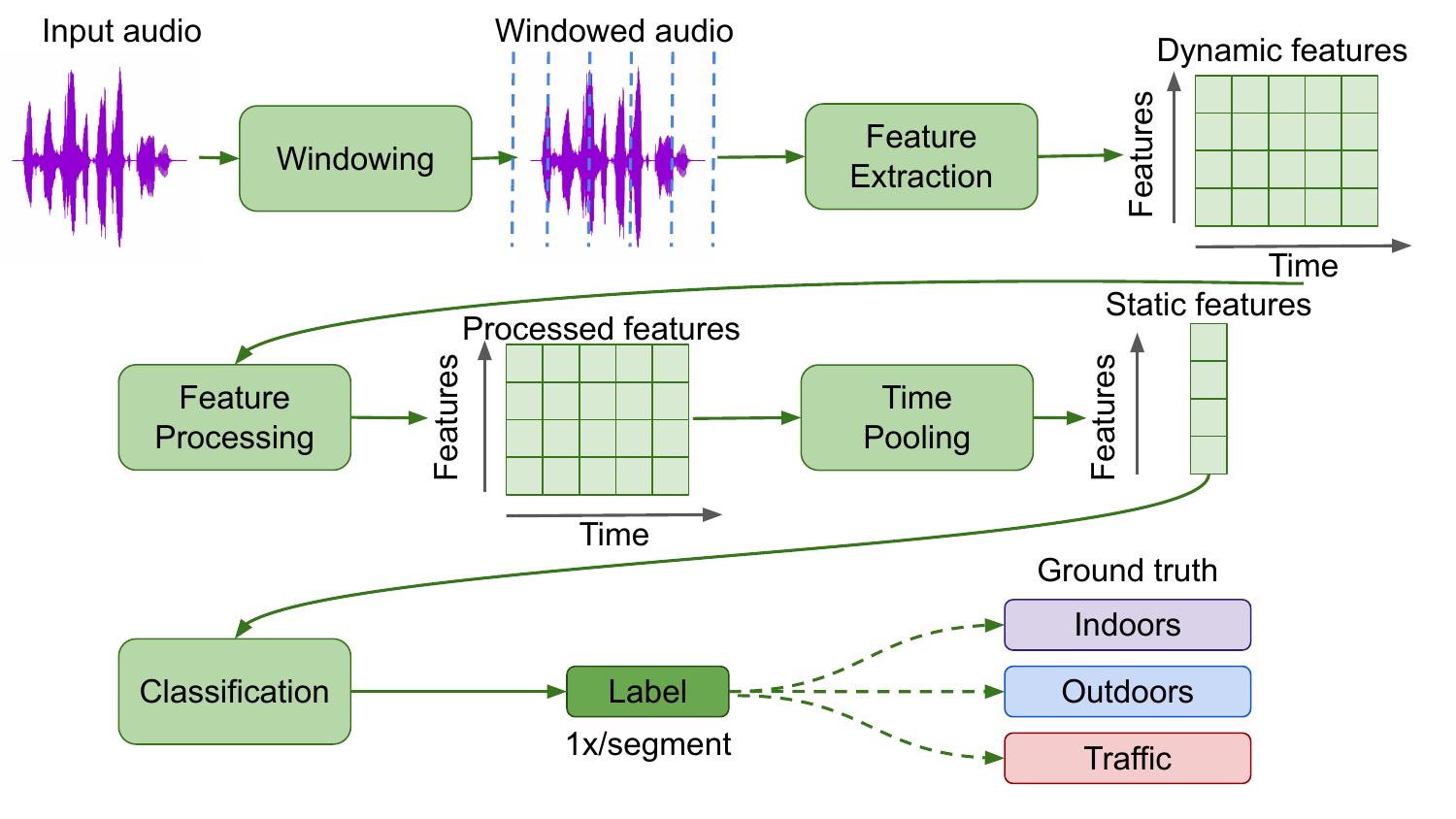}}~%
    \subfloat[Sound event detection]{\includegraphics[width=.48\textwidth]{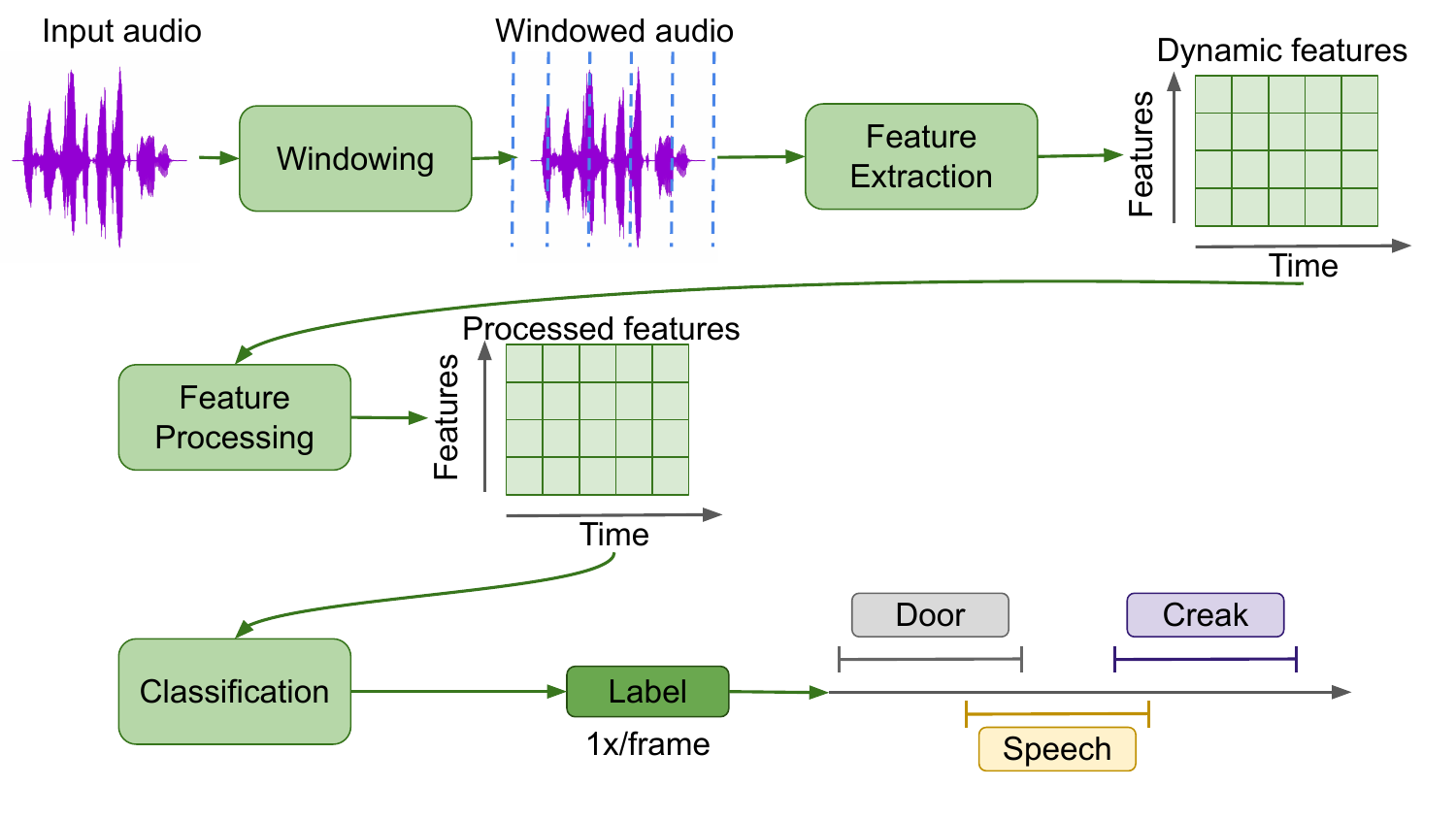}}
   \caption{
   Overview of supervised training using a traditional pipeline for \ac{ASC} (left) and \ac{SED} (right).
   Both pipelines begin with a windowing of the input audio signal (generally, in overlapping frames), followed by feature extraction performed on each frame to generate dynamic features.
   These dynamic features are optionally processed by another module, resulting in what we label `processed' features.
   In the case of \ac{ASC}, these features are `pooled' (aggregated) over their time dimension before being propagated to a final classification step.
   In contrast, this pooling step is omitted for \ac{SED} as it needs to generate one output per frame.
   Crucially, \underline{all the intermediate steps can be differentiable} and be implemented as part of a single \ac{DNN} architecture.
   }
   \label{fig:training}
\end{figure*}

The typical approach to sound event or acoustic scene classification is to use straightforward supervised training~\citep{Mesaros21-SED}.
A machine learning system, nowadays most often a \ac{DNN}, is trained using examples of audio and corresponding annotations, to learn a mapping between the acoustic information presented to it (in the form of acoustic features) and the target classes to be recognised.


Irrespective of the task, a system for supervised learning consists of the same pipeline.
The audio signal is typically transformed into a more compact representation through feature extraction, creating a representation matrix as input for the machine learning algorithm.
These features represent key information about the audio content in a form suitable for learning and can be obtained through hand-crafted processing (e.\,g., time-frequency representations like \ac{STFT} or mel energies) or through representation learning, in which case they are referred to as \emph{embeddings} (though this case is more relevant for transfer learning; see below).
In the latter case, a \ac{DNN} is trained separately or simultaneously with the subsequent classifier.


On the other hand, the annotation is encoded into a representation to be provided as target outputs in the learning process.
This encoding is task-specific and serves as a representation of the information with respect to the audio clip or segments that form the unit-of-analysis.
For example, \ac{ASC}, as a multi-class single-label classification task uses a one-hot encoding of the target classes, showing that one class is active in one clip; \ac{SED}, as a multi-class multi-label classification task, uses a time-dependent multi-hot encoding, showing that multiple event classes can be active in each analysis time frame.
Consequently, in the case of \ac{DNN}, the networks trained for these tasks have task-specific output layers, namely a softmax layer for the single-label and a sigmoid layer for the multi-label case.
The rest of the network architecture can be very similar, though in time-dependent outputs like SED or tracking it is beneficial to utilise some type of recurrent units able to model sequences. 


The entire process is illustrated in \cref{fig:training}, which lays out the different components of a standard audio analysis pipeline.
Given an audio signal as the input, this signal is initially \emph{windowed} (i.\,e, broken into smaller, overlapping chunks, often processed by a suitable windowing faction to avoid artefacts), and subsequently passed to a feature extraction process that extracts features for each frame.
These features are subsequently processed by another module, before being passed into a final classification component.
In the case of \ac{ASC}, the processed features are aggregated over time before being given to the classification layer; for \ac{SED}, in contrast, this time-aggregation step is omitted as the goal is to produce segment-level decisions.

Importantly, any or all of the steps shown in \cref{fig:training}, for both \ac{ASC} and \ac{SED}, may be learnable from data.
For example, our ``Feature Processing'' module is typically realised using convolution -- or, nowadays, attention -- layers~\citep{Kong20-PLP, Gong21-AAS}, whereas the ``Classification'' head is implemented as \acp{MLP}~\citep{Kong20-PLP, Gong21-AAS}.
``Feature extraction'' can be an additional preprocessing step, e.\,g., used to extract mel spectrograms~\citep{Kong20-PLP, Gong21-AAS}, or be included in the model as is the case in ``end-to-end'' models~\citep{Trigeorgis16-AFE}.
``Time Pooling'', too, may be learnt, as is the case for attention-based pooling\add{{~\citep{Zhao19-EDS}}}.
Further, in models that rely on learnable ``frontends'', like LEAF~\citep{Zeghidour21-LAL}, distinguishing between some initial layer(s) charged with learning feature extraction, and subsequent (convolution/attention) ones that process those features, might help understand the architecture better.
Nowadays, end-to-end models can also incorporate the ``Windowing'' step using learnable components~\citep{Liu22-LTS}.
In short, all different combinations of interlacing learnable and non-learnable components can be utilised.
However, it is important to note that -- at the other extreme -- all steps of our pipeline can, in principle, be fixed and designed by experts (i.\,e., the expert-based systems of yore which have been largely abandoned), though, in most cases, the ``Classification'' module is, at least, trained on data.

Earlier versions of classification systems relied indeed on such traditional classifiers.
For example, the \ac{DCASE} 2013 and 2016 challenges saw heavy use of \acp{SVM} and \ac{GMM}, with a transition to \ac{DL} after 2016.
Initially, prominent \ac{CNN} architectures commonly used in computer vision, such as ResNet~\citep{He16-DRL} or VGG~\citep{Simonyan15-VDC}, were employed to address challenges within the domain of audio and sound analysis.
This methodology involved the conversion of audio streams into spectrograms, thereby representing them visually.
These approaches lead to the development of specialised networks tailored to specific audio-related tasks.
However, as computational power increased, so did the employed networks, and the relatively limited size of employed datasets (compared, for instance, to computer vision) posed a problem.
This was solved by large-scale pretraining and finetuning -- as we see in the next section.


\subsection{Transfer learning}
\label{sec:tl}

A more recent line of approaches relies on transfer learning\add{{~\citep{Zhuang20-ACS}}}, i.\,e., training a network with a large amount of data somewhat related to the task, and finetuning it with the downstream task data, which is usually of smaller size.
\add{The key idea driving the adoption of transfer learning in {\acp{DNN}} is the transfer of useful representations to the target task -- also known as \emph{feature re-use}{~\citep{Neyshabur20-WIB}}.}
This approach entails training a model on a particular upstream task, such as audio tagging, for which large datasets are available, and subsequently training the model further on a specific downstream task, such as \ac{ASC}.
This process, known as \textit{finetuning}, adapts a pretrained model to a new task by leveraging the previously learnt weights, and thus exploits the capability of recognising relevant features from the initial training phase. 

The use of transfer learning increased substantially after the introduction of large datasets, like AudioSet~\citep{Gemmeke17-ASA}. 
This paved the way for the development of large-scale \acp{PANN} for audio pattern recognition \citep{Kong20-PLP}, which have been trained on AudioSet.
Through this training, the models learnt robust audio representations that generalise well to other tasks -- though task and acoustic similarity play an important role~\citep{Triantafyllopoulos21-TRO}.
Therefore, these models could successfully be adapted to the various audio analysis tasks we described in \cref{sec:tasks}.
Following their initial success, several other pretraining models emerged, like BEATs~\citep{Chen23-BAP} or \ac{AST}~\citep{Gong21-AAS}, which changed the pretraining task (e.\,g., from supervised audio tagging to self-supervised modelling), the model architecture, or both.
\add{As we discuss below, the emergence and widespread adoption of transfer learning constitutes the first step towards foundation models.}

\subsection{Audio captioning}
\label{ssec:captioning}

\begin{figure}[t]
    \centering
    \includegraphics[width=\columnwidth]{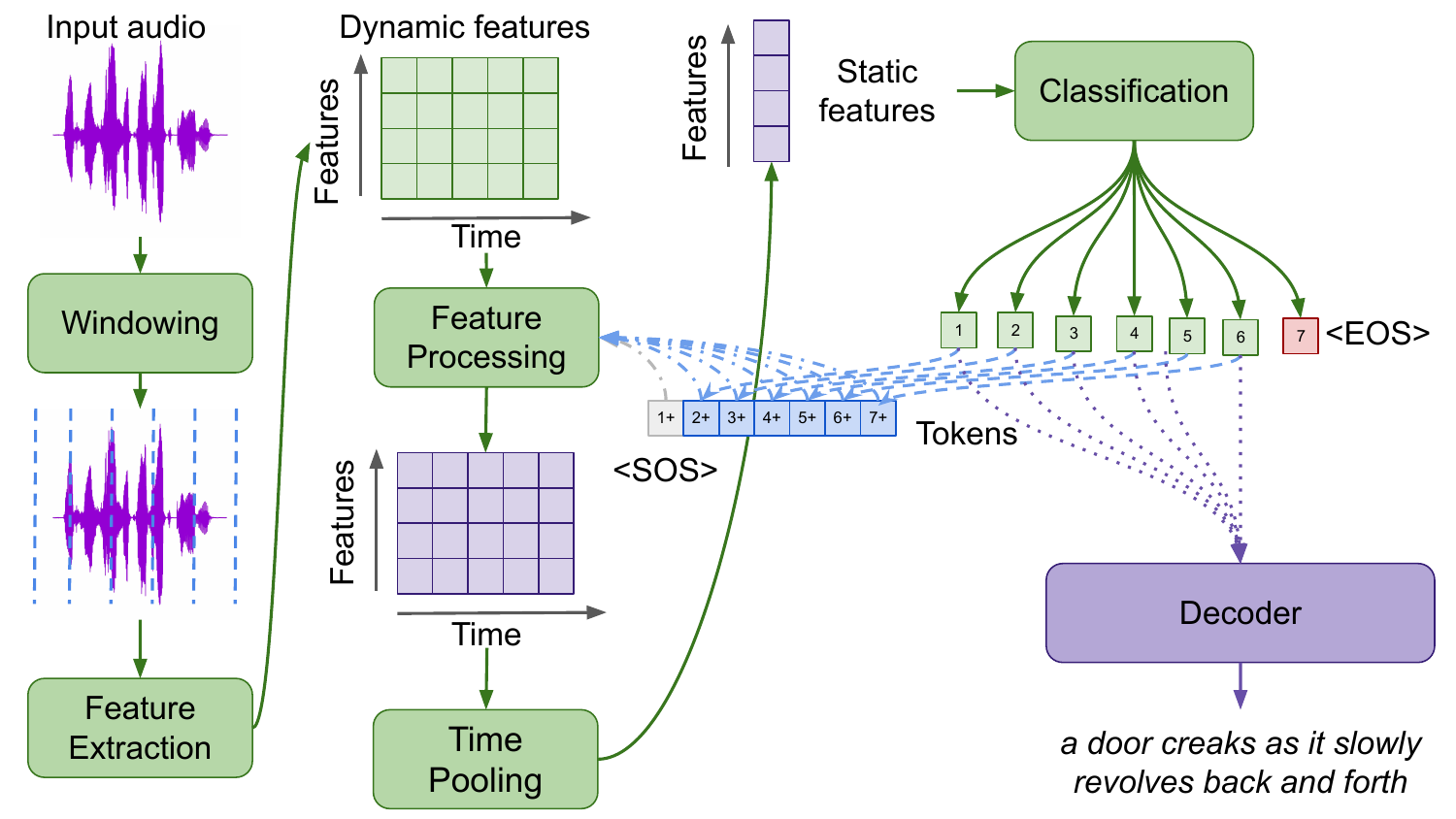}
   \caption{
   Overview of a `traditional', task-specific captioning pipeline.
   \remove{Similar to \mbox{\ac{ASC}} and \mbox{\ac{SED}} (see \mbox{\cref{fig:training}}), the pipeline comprises windowing, feature extraction and transformation, and a final classification step.}
   The main difference \add{to \mbox{\ac{ASC}} and \mbox{\ac{SED}}} is that the generation of output text tokens happens \textbf{autoregressively}.
   Beginning with an \ac{SOS} token, the model starts to output the next token one-by-one.
   Each new token is sequentially added to the list of tokens that become the input to the model in order to generate the next token in the series.
   The process repeats until an \ac{EOS} token is predicted\remove{ (or a maximum number of tokens is used)}.
   The final sequence\remove{ (barring \mbox{\ac{SOS}} and \mbox{\ac{EOS}})} is `decoded' back to strings to produce the final output.
   \remove{The tokens can be injected to the model in all possible places (input, dynamic features, processed features, static features); however, it is most common to add it to an intermediate step (also named `decoding'); this is what we show here; see text for more details.}
   }
   \label{fig:captioning}
\end{figure}

We single out \acl{AC} as that task is solved on slightly different principles than \change{others,}{those tasks} which are amenable to standard supervised learning~\citep{Drossos20-CAA, Mei24-WAC}.
\cref{fig:captioning} shows an overview of the standard pipeline.
Specifically, \ac{AC} systems must generate sequences of words, which are not temporally aligned with the input, and the number of which is variable.
In essence, it is a sequence-to-sequence task with non-synchronised labels.
Unlike a task like \ac{SED}, where each label in the model output sequence or annotation corresponds to a certain temporal segment in the input audio, one word in a caption does not necessarily correspond to a specific temporal segment in the input audio, and the length of the output sequence does not deterministically depend on the length of the input.
This is because the description of an auditory scene partially depends on its context -- an hour-long silence can be described with a single word, whereas a few seconds in a busy city centre might require a paragraph.
This introduces an additional problem for \ac{AC} models -- that of knowing when to stop.

In general, this is handled by \emph{autoregressive} methods, which sequentially generate outputs until they produce an \ac{EOS} token.
Beginning with the \ac{SOS} token, each step produces a single token, which is then appended to the list of previous tokens and used in all subsequent steps.
After the \ac{EOS} token is predicted (or a maximum number of generation steps is finished), the list of tokens is converted to the resulting string (and, optionally, further corrected in a post-processing step which we omit here).

\begin{figure*}[t]
    \centering
    \begin{tabular}{c|c}
        \subfloat[Few-shot learning]{\includegraphics[width=.48\textwidth]{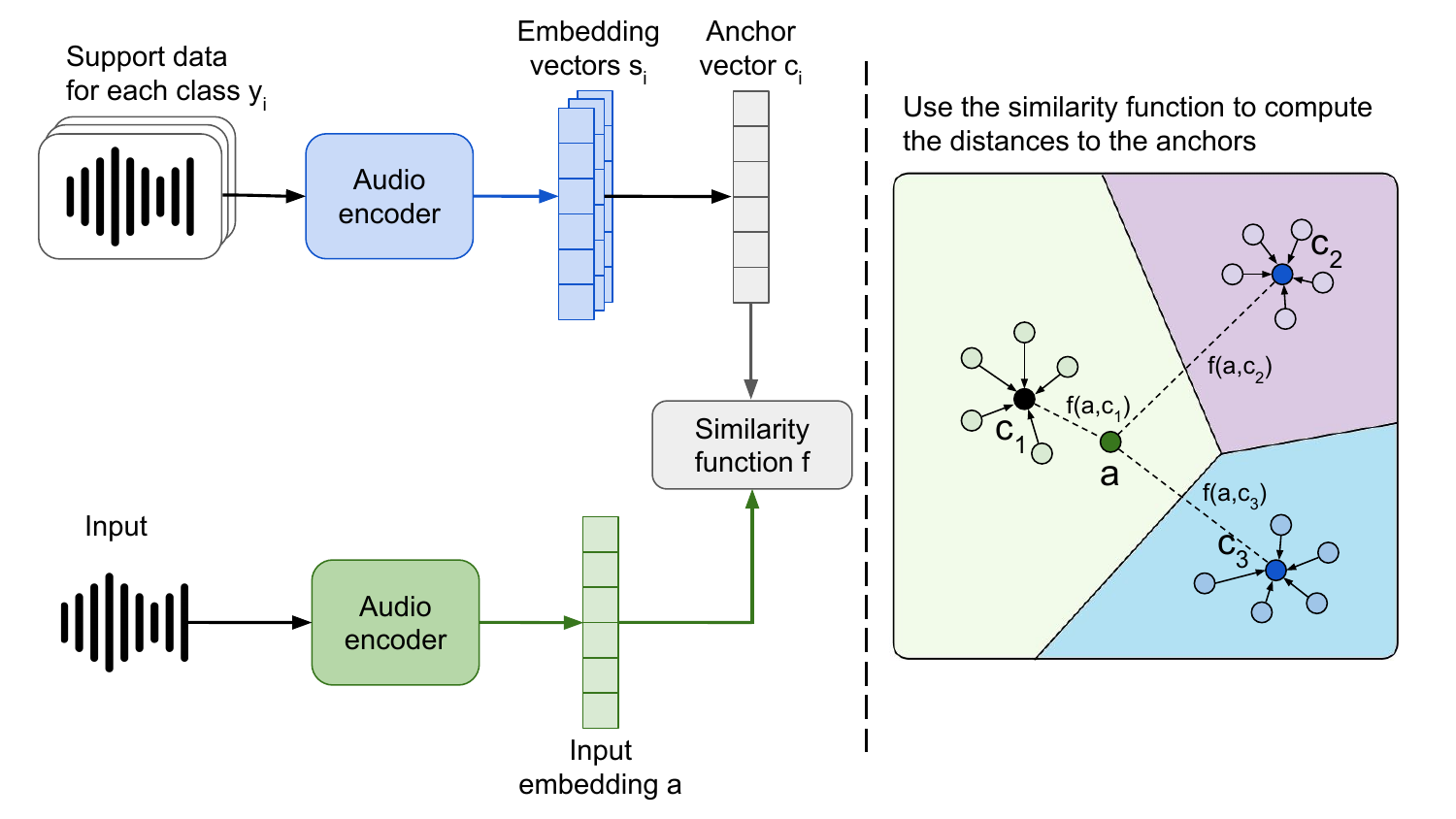}} &
        \subfloat[Zero-shot learning]{\includegraphics[width=.48\textwidth]{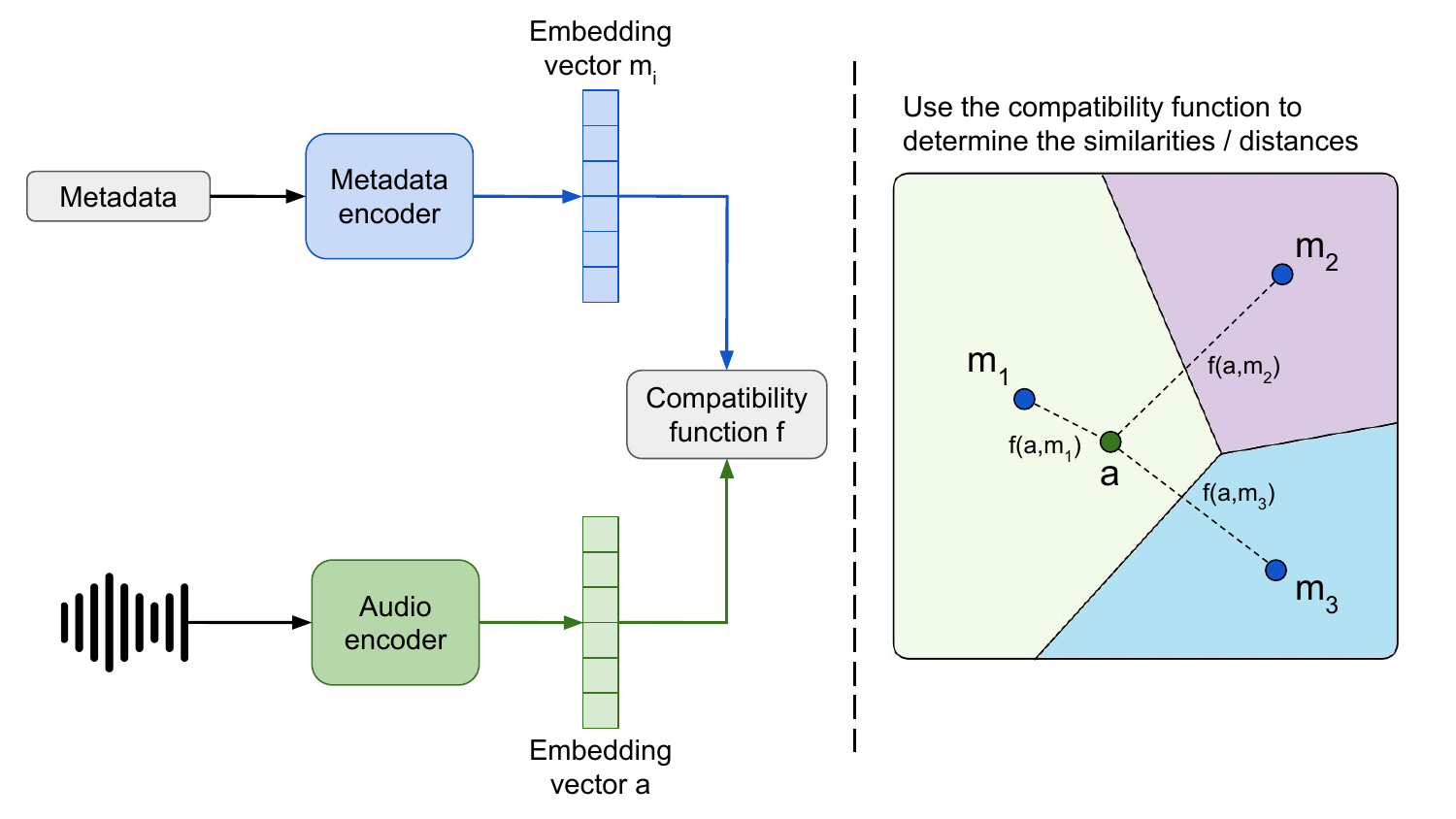}}  \\
    \end{tabular}
   \caption{
   Overview of the basic principles behind few-shot and zero-shot learning.
   \textbf{Few-shot learning (left):} The support data for each class $y_i$ are fed to an audio encoder, extracting the corresponding embedding vectors. These vectors are then merged (e.\,g., by taking the mean) to form the  anchor $c_i$ of class $y_i$. The input audio is fed to another or the same audio encoder to obtain an embedding vector $a$. Then, a similarity function $f$ is applied to $a$ and all anchors $c_i$ to measure the distances. The class of the anchor closest to $a$ is considered the correct class.
   \textbf{Zero-shot learning (right):} The metadata (e.\,g., text descriptions) for each class $y_i$ are fed to an encoder which extracts an embedding vector $m_i$ for each class $y_i$. The embedding vector $a$ of the input audio is obtained by feeding it to an audio encoder. The audio input is considered to belong to the class $y_i$ whose embedding vector $m_i$ has the highest similarity (or closest distance) to $a$.
   }
   \label{fig:fsl-zsl}
\end{figure*}

Previous tokens can be injected into the classification model in all possible positions: at the input stage (e.\,g., concatenated to the audio waveform/features), at an intermediate state (i.\,e., used to condition the generation of processed features), or closer to the output (e.\,g., by being concatenated to the static features).
All these approaches are viable ways to add previous information to the model and generate the next token autoregressively.
In \cref{fig:captioning}, we show the most standard approach, which is that of injecting the text tokens in some intermediate state.
Note that some works will denote the part of the architectures that takes as input the intermediate audio representations and previous tokens as the ``decoder''~\citep{Drossos20-CAA}, as that is the step which takes care of generating a new token.
In our case, we opted to use this term for the inverse mapping from tokens to strings -- this allows us to keep the same terminology for our model by referring to its output part as the classification module.
This remains consistent with our diagram for \ac{ASC} and \ac{SED} (\cref{fig:training}), as, ultimately, the process of predicting the next token given the previous ones is merely one of classification.

Finally, we point out that one key feature of \ac{AC} models is that, by necessity, they rely on language models to produce the output sequence.
While this makes \ac{AC} a prime target for audio \acp{FM}, not all \ac{AC} models \emph{are} \acp{FM}.
Indeed, the baseline for Clotho, which was the first dataset for \ac{AC}, used a shallow language model comprising a simple \ac{LSTM}~\citep{Drossos20-CAA}.
Nevertheless, as we see below, \ac{AC} is one of the key tasks used for training and evaluating audio \acp{FM}.

\subsection{Few-/Zero-shot learning}
\label{sec:zsl}

We close with a brief discussion of few- and zero-shot learning for more traditional methods.
Even though both of those are two of the most heavily advertised capabilities of \acp{FM}, researchers have been pursuing them for a long time, and certainly before the introduction of \acp{FM}. The basic principles of both are depicted in \cref{fig:fsl-zsl}.

\textbf{Few-shot learning:} Few-shot learning corresponds to the classification of a sample given a small ``anchor'' or ``support'' set for each possible category.
This allows for the introduction of new classes during test-time, a highly-desirable feature when working on problems where labelled data is sparse, such as ecoacoustics~\citep{Morfi21-FBE, Stowell22-CBW}.

The goal of a standard \ac{ML} system is to predict targets from inputs.
Few-shot learning introduces an intermediate step: the system must learn to associate inputs to anchors from different targets and use that association to make its decision (e.\,g., by picking the closest target according to a suitable distance function).
Oftentimes, anchors are referred to as \emph{prototypical} instances~\citep{Snell17-PNF} (though there is no requirement that these instances be prototypical in the colloquial sense).

Standard methods for few-shot learning include Siamese~\citep{Bromley93-SVU} and prototypical networks~\citep{Snell17-PNF, Liang24-MTD}.
Both map inputs and anchors to an embedding space using an encoder (which may or may not be shared).
Then, a \emph{compatibility function} computes a notion of similarity between the embeddings.
The final prediction is made according to the compatibility of the input with each set of anchors; a `hard' rule is to assign the class with the maximum compatibility as the prediction, while a `soft' rule takes into account the compatibility to all anchors for propagating the label.

\textbf{Zero-shot learning:} Zero-shot learning is the most extreme version of few-shot learning~\citep{Xie21-ZAC, Dogan22-ZAC, Elizalde23-CLA, Gebhard24-EMI}.
In principle, a zero-shot learning workflow is almost identical to a few-shot learning one: a compatibility function is used to map the learnt representations of audio inputs to (other learnt representations of) categories.
Here, however, there are no \emph{audio} data available for the categories under consideration.
Instead, the model needs to learn to associate audio inputs with non-audio metadata.
This metadata may be textual descriptions, visual imagery, or numeric/binary attributes of the target audio classes. 
Zero-shot learning models operate typically by projecting audio and metadata into the same shared space, and classifying an audio sample to the metadata class that is closest to it in the shared space.

\section{Audio foundation models}
\label{sec:afm}

Recent years have seen the introduction of so-called \acp{FM} in various areas of \ac{AI}.
\citet{Bommasani21-OTO} defined them as follows: ``[...] [foundation] models [...] trained on broad data (generally using self-supervision at scale) that can be adapted to a wide range of downstream tasks''.
As such, they are general-purpose models that can be co-opted for a variety of tasks.
This is in contrast to `old-school' \acp{DNN} that were trained to be \emph{specialists} in particular tasks.
In this section, we review their inner workings, training processes, and most recent iterations, starting with a definition of what constitutes a foundation model.

\subsection{A definition of foundation models}
\label{sec:definition}
The definition of \acp{FM} is rather vague.
Specifically, it raises the question of whether \acp{DNN} trained in supervised fashion for a particular task, but later shown to be effective on a variety of downstream tasks, also constitute foundation models.
Examples of such very successful models are:
\acp{PANN}~\citep{Kong20-PLP}, which were trained on AudioSet to predict sound events but later found effective on a broad range of problems;
\wav~\citep{Baevski20-W2A}, which was trained on a self-supervised task on LibriSpeech but then found to generalise to a wide gamut of speech-related tasks~\citep{Wagner23-DOT};
\remove{or }\emph{Whisper}~\citep{Radford23-RSR}, which was trained on a large, multilingual \ac{ASR} corpus but successfully finetuned even on such disparate tasks like audio tagging~\citep{Gong23-WNA}
\add{; or \emph{neural audio codecs} which can efficiently compress audio, such as \emph{EnCodec}{~\citep{Defossez23-HFN}}, \emph{SoundStream}{~\citep{Zeghidour21-SAE}}, or \emph{SemantiCodec}{~\citep{Liu24-SAU}}}.
These models all learn very powerful, generalisable representations during their initial training which can be exploited for downstream tasks using finetuning.

We turn again to \citet{Bommasani21-OTO} for a clarification (emphasis is ours):
``Though foundation models are based on standard deep learning and transfer learning, their scale results in new \emph{emergent} capabilities, and their effectiveness across so many tasks incentivises \emph{homogenisation}.''
Therefore, according to their original definition, \acp{FM} need to exhibit \emph{emergence} and \emph{homogenisation}.
We examine these terms in the following two sub-sections.

\subsubsection{Homogenisation}
Homogenisation reflects the broader tendency in \ac{DL} to rely on a common set of architectures, often initialised with a pretrained state, for a wide variety of tasks.
The standard example is ImageNet~\citep{Deng09-IAL}.
Following its introduction, a wave of \ac{AI} advances followed that utilised it as their starting point, many of which relied on using the same architectures, with weights initialised by pretraining on ImageNet or other large datasets, and then finetuning them on the target downstream task.
The same tendency was shown in \ac{NLP} with the introduction of \ac{BERT}~\citep{Devlin18-BPO} and its successors, and later in audio with \acp{PANN}~\citep{Kong20-PLP}, \wav~\citep{Baevski20-W2A}, and others. 
From this perspective, \acp{DNN} with a widespread use and an `aptitude' for different tasks (\emph{importantly:} after finetuning) do constitute foundation models.

\subsubsection{Emergence}
Emergence corresponds to the `sudden' appearance of new capabilities (that the model was not explicitly trained for) as an aftermath of scale~\citep{Bommasani21-OTO}.
This is typically observed in \acp{LLM}, which -- after a specific data and complexity limit is overcome -- showcase the ability for \emph{in-context learning}~\citep{Dong22-ASO}\add{, a particular style of learning which arises from a description of the target task during inference time}.
A standard example is \emph{GPT-3}~\citep{Brown20-LMA}, which had 175 billion parameters compared to the 1.5 billion of its predecessor (\emph{GPT-2}~\citep{Radford19-LMA}).
The larger model (trained on more data) could be adapted to novel downstream tasks with the introduction of a \emph{prompt} -- in general, a prefix that describes the required task (see below).
This new, `emergent' property was not introduced by a change in the training algorithm, but rather as a byproduct of scale.
Importantly, prompting allows the use of \acp{FM} \emph{without task-specific finetuning}, a dramatic deviation from the previous state-of-the-art which relied on `traditional' transfer learning (by means of changing model weights using an adaptation dataset).
This opened up exciting new avenues for the utilisation of \acp{FM} as a singular, consolidated point-of-entry for building new applications.
Seen in this light, earlier \ac{DNN} `backbones' which had to be finetuned on the target task, like \acp{PANN} or \wav, do \emph{not} constitute \acp{FM}.

\subsubsection{\add{What is a foundation?}}

\add{
We note here a contradiction in the work of \mbox{\citet{Bommasani21-OTO}}.
The requirement to seamlessly handle different tasks \emph{without finetuning}, i.\,e., without changing a model's weights, is part of their definition of {\acp{FM}}.
This calls for a flexibility in their output space, which should be able to accommodate multiple tasks, even tasks that have not been seen during training.
This is a crucial difference to traditional pipelines, as {\acp{FM}} must be \emph{open-ended}, whereas traditional models need only work for a constrained set of tasks.
To achieve this, contemporary {\acp{FM}} rely on \emph{human language} as their output modality{~\citep{Bommasani21-OTO}} (usually in the form of text, but increasingly also using speech).
In a sense, they resemble the audio captioning models presented in \mbox{\cref{ssec:captioning}}. 
However, the scope of application is much broader than the one simpler audio captioning models introduce, as they operate within a single dataset (for instance, the baseline model by \mbox{\citet{Drossos20-CAA}} has a very small fixed vocabulary, capturing only the words present in the Clotho dataset). 

In another part of their work, however, \mbox{\citet{Bommasani21-OTO}} expand on the term \emph{foundation} as follows:
``In particular, the word `foundation' specifies the role these models
play: a foundation model is itself incomplete but serves as the common basis from which many task-specific models are built via \emph{adaptation}'' (emphasis ours).
This adaptation can be done in the form of prompting or finetuning and thus reopens the door to reconsider unimodal backbones as {\acp{FM}}.

\textbf{Therefore, we decided to adopt a balanced perspective by designating two \emph{tiers} of audio \mbox{\acp{FM}}.}
\emph{Tier-1} \mbox{\acp{FM}} are unimodal, audio-only models that have found widespread adoption following fine-tuning.
\emph{Tier-2} \mbox{\acp{FM}} additionally incorporate \mbox{\acp{LLM}} as one of their unimodal components -- at the \emph{decoding} stage.
This endows them with the flexible output space which is a prerequisite for \emph{emergence without finetuning}.
}

\subsection{\add{\emph{Tier-1} audio foundation models}}
\label{sec:tier1}
\add{
In the present section, we discuss \emph{Tier-1} audio {\acp{FM}}.
These are unimodal models which take as input audio and output a representation of it.
Crucially, this representation should generalise across multiple tasks.
There are three major distinguishing factors for differentiating between models: their \emph{architecture}, the \emph{data}, and the \emph{task} they have been trained on.
We expand on the different archictures and training tasks below.
The different datasets used for training are instead presented in \mbox{\cref{ssec:datasets}}.
}

\add{
\textbf{Architectures:}
Development in audio architectures closely mirrored advances in the related fields of computer vision and {\ac{NLP}} (another sign of homogenisation~\mbox{\citep{Bommasani21-OTO}}).
The first \emph{Tier-1} audio {\acp{FM}} were \emph{convolutional} in nature.
Among them, it is the family of \emph{PANNS} which stands out~\mbox{\citep{Kong20-PLP}}, with the most widely-used models being \emph{CNN14} and \emph{CNN10}.
Both co-opt the \emph{VGG} architecture for audio~\mbox{\citep{Simonyan15-VDC}}.
A simpler model was instead chosen by \mbox{\citet{Niizumi22-BFA}} for their work on \emph{BYOL-A}.
Convolution also formed the backbone of early \emph{neural codec} models, like \emph{EnCodec}~\mbox{\citep{Defossez23-HFN}} (see below).
However, following the introduction of \emph{attention}~\mbox{\citep{Vaswani2017-AIA}}, this particular model component was quickly adopted for audio as well.
\emph{AST}~\mbox{\citep{Gong21-AAS}}, \emph{AudioMAE}~\mbox{\citep{Huang22-MAT}}, \emph{wav2vec2.0}~\mbox{\citep{Baevski20-W2A}}, \emph{WavLM}~\mbox{\citep{Chen22-WLS}}, and \emph{Whisper}~\mbox{\citep{Radford23-RSR}} are all examples of models that rely on self-attention.
}

\add{
\textbf{Tasks:} While model architectures already show evidence of homogenisation, especially after the rise of \emph{transformers}, pretraining tasks show a much larger variability.
The simplest approach is to train a model in supervised fashion.
This was followed by \mbox{\citet{Kong20-PLP}} for \emph{PANNS} by training them for audio tagging on AudioSet.
Similarly, \mbox{\citet{Radford23-RSR}} trained \emph{Whisper} for \mbox{\ac{ASR}} on a large (private) speech dataset.
}

\add{
The next `usual suspect' for pretraining is reconstruction.
In the simplest possible case, these are autoencoders which aim to reconstruct the input.
Neural codecs also follow this principle~\mbox{\citep{Defossez23-HFN}}.
Importantly, though, their goal is to \emph{compress} the audio input while allowing for (near-)perfect reconstruction.
Thus, while they are trained on an autoencoding loss, they typically feature an intermediate quantisation step (such as residual vector quantisation~\mbox{\citep{Barnes96-AIR}}).
This focus on learning more efficient, shorter codes for the audio input make them highly appropriate as audio encoders for \emph{Tier-2} {\acp{FM}} as it allows them to side-step the limited context length of {\ac{LLM}} decoders~\mbox{\citep{Hsieh24-HWT}}.
}

\add{
There is also a plethora of methods following a {\ac{SSL}} paradigm.
These methods manipulate the input data itself to create proxy tasks.
Oftentimes, parts of the input are corrupted with noise, or \emph{masked}, i.\,e., a variant of the standard autoencoding task~\mbox{\citep{Huang22-MAT}}.
The masking can also be done in the space of intermediate representations, as for \emph{wav2vec2.0}~\mbox{\citep{Baevski20-W2A}}.
Alternatively, clustering into discrete units can be used to create pseudo-targets for pretraining, as done for \emph{HuBERT} in the context of speech~\mbox{\citep{Hsu21-HSS}}.
Moreover, the reconstruction can be done in a causal fashion, though this is mostly pursued in the context of speech and is inspired by {\ac{NLP}} research~\mbox{\citep{Chen22-WLS}}.
}

\add{
Finally, a widely-used form of {\ac{SSL}} is \emph{contrastive learning}.
This relies on pulling closer the representations of instances that should be similar while pulling further apart those that should be different.
These associations are determined by the approach.
For instance, \emph{BYOL-A} pulls closer augmented views of the same audio input while pushing apart views of different instances~\mbox{\citep{Niizumi22-BFA}}.
Contrastive learning can also be combined with the above, as in the case of \emph{wav2vec2.0}, which uses a contrastive loss to predict its masked intermediate representations. 
}

\add{
We note that we already mentioned \emph{transfer learning} as part of the conventional methods in \mbox{\cref{sec:tl}}
As in the case of audio captioning, the boundaries between conventional and foundation models are fluid.
A key component to differentiate {\acp{FM}} from `conventional' models that rely on transfer learning is \emph{scale}~\mbox{\citep{Bommasani21-OTO}}.
However, as is often the case with scale, it is eventually a matter of `degree'.
We are not very strict with our taxonomy as there are several factors which influence the adoption of a model as a `foundation'.
Among others, this is the generalisability of its representations, its ease-of-use, and the availability of checkpoints.
In other words, it is a `community decision', whereby models see their adoption change over time.
}

\add{
We end this section with a more in-depth discussion of \emph{CLAP}~\mbox{\citep{Elizalde23-CLA}}.
While we already mentioned it as a zero-shot method in \mbox{\cref{sec:zsl}}, there is a particular facet of \emph{CLAP} which sets it apart from other zero-shot methods -- the fact that it has been trained to generate audio-language mappings.
As such, it is a contrastive method like the ones discussed before, though not a self-supervised one as it relies on the presence of labels.
However, it features a similar flexibility in its output space as the one we defined as the key differentiating factor between \emph{Tier-1} and \emph{Tier-2} models.
A user can query it using natural language, and thus the model can provide `answers' to a very wide gamut of tasks without finetuning.
Nevertheless, we consider the fact that a human needs to provide specific queries a serious limitation compared to \emph{Tier-2} models which are autoregressive and can thus generate outputs autonomously.
As a result, we decided to include \emph{CLAP} here, rather than in the next section.
We acknowledge, though, its importance as an intermediate milestone between \emph{Tier-1} and \emph{Tier-2} audio {\acp{FM}} -- perhaps enough to position it in a midway \emph{Tier-1.5}.
}

\subsection{\add{\emph{Tier-2} audio foundation models}}
\label{sec:tier2}

\add{
We now turn to the more advanced, \emph{Tier-2} family of {\acp{FM}}.
These are more flexible than \emph{Tier-1} models in that they provide their output in natural language.
They additionally allow for text inputs which help condition the model to the requirements of the user.
Taken together, these two functionalities facilitate the \emph{emergence} of capabilities that have not been trained for \emph{without finetuning}.
This is a major leap beyond the previous generation of \emph{Tier-1} {\acp{FM}} and hinges on the successful coupling of audio and language components.
The audio components are often, but not always, the \emph{Tier-1} audio {\acp{FM}} we discussed above.
As we see below, in the context of \emph{Tier-2} models they are utilised as \emph{audio encoders} that are coupled with an additional \emph{text encoder} before their outputs are passed to a \emph{text decoder}.
We thus begin by discussing how text components are trained before continuing with how audio-language interfaces are created.
}

\subsubsection{\change{Preliminaries}{Autoregressive models}}
\label{ssec:preliminaries}
Before we proceed with our overview of \change{foundation models}{\emph{Tier-2} {\acp{FM}}} and how they can be used for audio tasks, we introduce a set of preliminaries.
\add{As mentioned in \mbox{\cref{sec:definition}}, we consider the output space of \emph{Tier-2} {\acp{FM}} to be human language, i.\,e., \emph{text}. This means that the last stage of such {\acp{FM}}, the \emph{decoder}, is \emph{autoregressive} in nature -- it create sequences of outputs element-by-element.}
\remove{For our purposes, we only consider \emph{autoregressive} models -- models which can create sequences of outputs on an element-by-element nature.}
This captures most \change{foundation models}{\emph{Tier-2} {\acp{FM}}} that are available today\add{ (see \mbox{\cref{ssec:existing}})}.

Specifically, these models are tasked with generating a sequence of outputs $S = (\pmb{x}_1, ..., \pmb{x}_c)$, where $c$ is the \emph{context length} supported by a particular architecture and $\pmb{x}_{*}$ are multidimensional vectors.
In the context of \acp{LLM}, $\pmb{x}_{*}$ are typically referred to as ``tokens''\footnote{While there is not a direct mapping to audio or vision data, we will nevertheless stick with this term as it is widely encountered in the literature.
} -- vectors representing subword units.
Once predicted, tokens can be decoded back to a textual representation that is presented to the user.
This formulation allows for extensive generalisation capabilities, as contemporary \acp{LLM} can process inputs with hundreds of thousands, or even millions, of tokens.
\add{Similarly, in the case of \mbox{\acp{FM}}, tokens include both the subword units of the task description and the encoded audio units of the audio input.}

Seen in this light, a \change{foundation model}{\emph{Tier-2} {\ac{FM}}} is a function $f(\cdot)$ which approximates a probability distribution $p(\pmb{x}_{n}|{\pmb{x}_1},\pmb{x}_2...,\pmb{x}_{n-1})$.
$f(\cdot)$ must be evaluated at each step of the training process.
Sampling, or decoding\footnote{Here, as is often done in the literature, we use decoding to denote the selection of the best candidate $\pmb{x}_n$, \emph{not} the inverse transformation of tokens into text.}, can then proceed depending on the strategy of choice (e.\,g., greedy decoding will select the token with the highest probability).
Sampling stops either once a fixed output length is reached or when an \ac{EOS} token is predicted.

This autoregressive nature of most foundation models allows for a lot of flexibility on the types of inputs and outputs that a model can handle.
On the input side, tokens represent subword units for text-only \acp{FM}; for multimodal \acp{FM}, tokens also encode other kinds of inputs.
Moreover, the generation of tokens can be initiated from a particular sequence (what is generally referred to as \emph{prompting}).
In this case, the \ac{FM} will condition its generation process on those given inputs, and thus procure an output sequence that matches the intent of the input.

This modularity, coupled with the malleability of tokens on the output side, allows these models to tackle a variety of tasks.
While they are initially trained to model (text) sequences, they can also be used for prediction tasks by specifying a suitable prompt.
For example, an \ac{LLM} might be faced with a yes/no question (e.\,g., ``Is the colour of the sun yellow?'') and its answer interpreted as the prediction to the particular task.
As for the input sequences, the output tokens can also be interpreted as something other than language (e.\,g., as audio or images).
This allows us to cast virtually any task as autoregressive and `prompt' a\change{ foundation model}{n {\ac{FM}}} to solve it.

In the case of audio analysis, we are dealing exclusively with the description of acoustic properties.
This requires us to make the audio to be described a part of the input.
The output, however, can be restricted to come in the form of natural language, or even to specific words (e.\,g., the classes in an \ac{ASC} ontology).

\remove{
Audio {\acp{FM}} are typically created by connecting existing unimodal (audio and language) components (see \mbox{\cref{ssec:interfaces}}) -- with native multimodal support so far being the exception (\mbox{\cref{ssec:native}}).
The primary reason for this is the amount of computational resources needed to train the individual components; as most research groups cannot replicate that, they rely on existing models (primarily {\acp{LLM}}) and extend their capabilities by attaching an audio module to them.
This is why we first discuss how these unimodal components are created in this section, before continuing to how they are connected in the next one.}

\subsubsection{Pretraining \change{unimodal foundation models}{text decoders}}
\label{ssec:pretraining}


As mentioned, the success of \ac{FM} largely depends on them consuming large volumes of unlabelled data.
\add{The same holds true for the text decoders of \emph{Tier-2} {\acp{FM}}.}
This is achieved by \emph{pretraining} them on data using tasks that do not require manual labelling.
Removing the dependence on labels allows to vastly increase the quantity and diversity of data used, and, consequently, the knowledge that an \ac{FM} can acquire during pretraining.
The pretraining stage, thus, involves the following two steps:
Pretraining \add{of text decoders} often takes the form of \ac{SSL}, where the model is trained on ``proxy tasks'' that do not require human-annotated or other `high-level' labels~\citep{Liu22-ASL}.
Instead, proxy tasks are generally designed to predict some part of an instance using other parts of the same instance -- e.\,g., by masking the part to be predicted (following the paradigm of \emph{BERT}~\citep{Devlin18-BPO}\remove{ for text and \mbox{\wav~\citep{Baevski20-W2A}} for audio}) or predicting future elements in a sequence from past elements in the same sequence~\citep{Radford19-LMA}.
Those tasks aim to force models to (implicitly) learn the underlying data-generating distribution; this is achieved because predicting missing parts of the input implicitly relies on knowledge about the data generation process~\citep{Zbontar21-BTS}.
In this case, the model receives a sequence of tokens and autoregressively predicts the next token based on the preceding ones by maximising the likelihood of the predicted tokens.\remove{ as in \mbox{\cref{eq:loss_function}}.}
The method helps the model learn various data tasks by predicting the next tokens in a sequence, thereby gaining an understanding of the data structure and context.
In this type of \ac{SSL} task, the log-likelihood of the predicted token is used as the loss function.
Another \ac{SSL} method depends on masking some tokens in a sequence and training the model to predict the masked tokens based on the surrounding context.
This helps the model learn the relationships between different parts of the data. 



\subsubsection{Language-audio interfaces}
\label{ssec:interfaces}

\begin{figure*}[t]
    \centering
    \begin{tabular}{c|c}
        \subfloat[Early fusion]{\includegraphics[width=.48\textwidth]{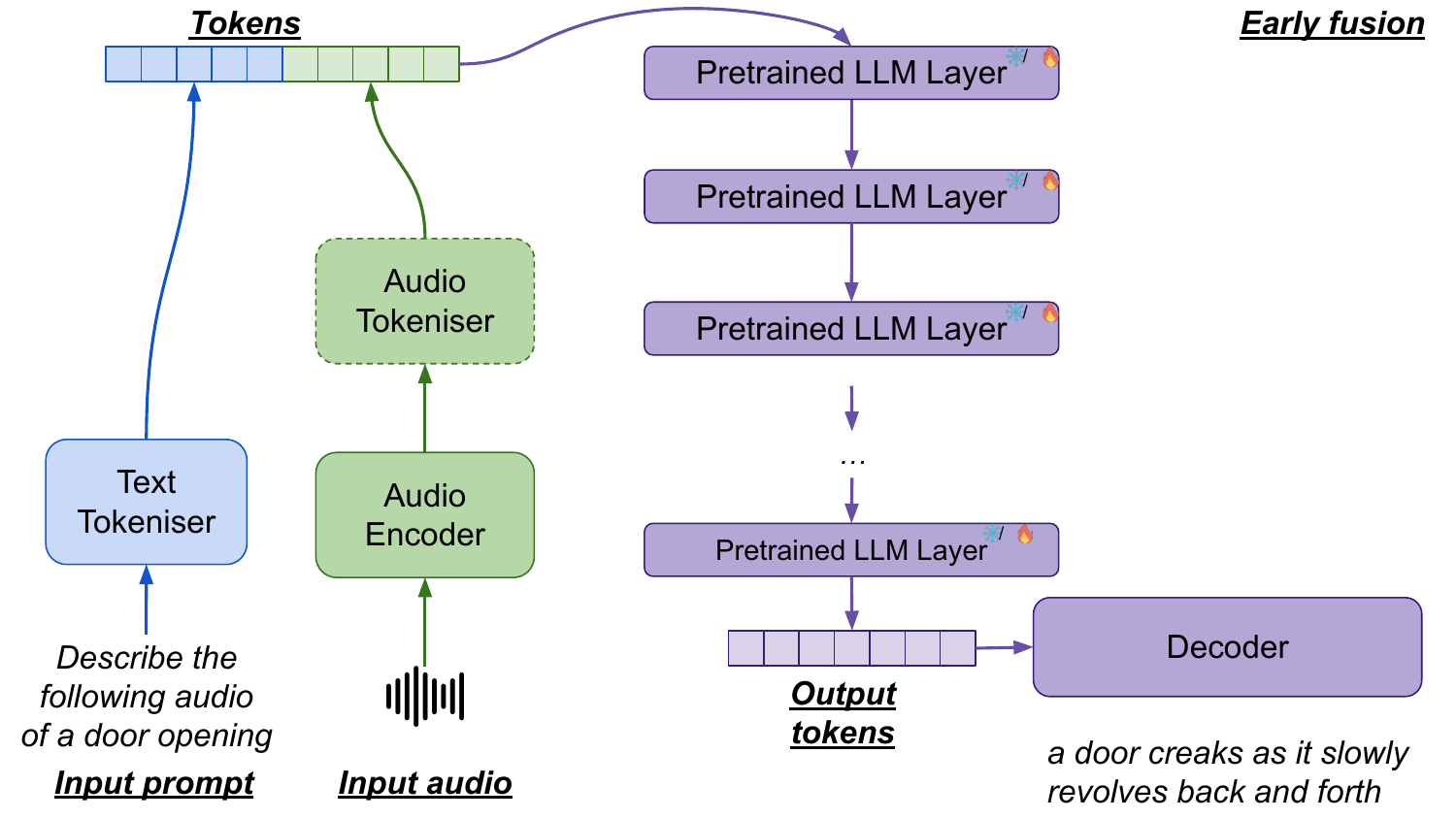}} &
        \subfloat[Deep fusion]{\includegraphics[width=.48\textwidth]{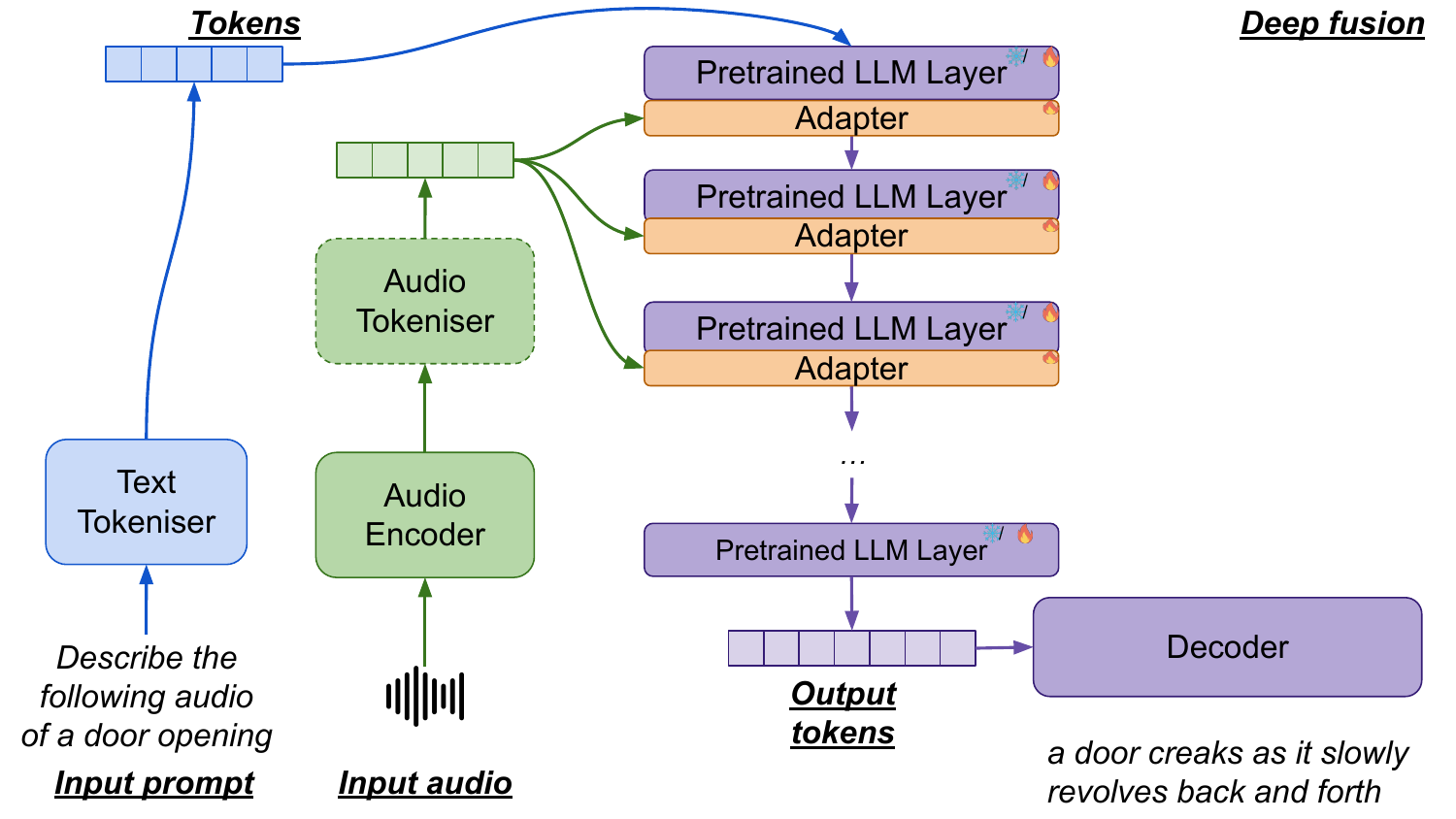}} \\
    \end{tabular}
    \caption{
    Overview of two main types of audio foundation model pipelines that combine audio with linguistic prompts.
    The input prompt is passed through a text encoder and tokeniser to obtain text tokens; this is used to provide context and instructions to the model.
    Accordingly, the audio input is passed through an audio encoder and tokeniser (the tokeniser may be unique to audio or the same as used for text or even skipped altogether).
    In the case of \textbf{early fusion} (left), these are concatenated with the text tokens, and subsequently passed to a pretrained large language model -- which may or may not be finetuned on additional data.
    In the case of \textbf{deep fusion} (right), the audio tokens are not concatenated (or interlaced) with the text tokens, but rather used to condition a series of trainable adapters which are used to modify the output of one or more (usually all) intermediate layers of an \ac{LLM}.
    In this case, the audio information is injected in the intermediate layers rather than the input.
    }
    \label{fig:foundation}
\end{figure*}

As we saw above, \emph{emergence} is a key attribute of \acp{FM}, with \textit{in-context learning}~\citep{Brown20-LMA} playing a fundamental role in it.
This type of learning relies on \emph{prompts} -- linguistic instructions which describe the task that has to be performed.
In order to achieve this for audio \acp{FM}, an interface between an audio \ac{FM} and a language \ac{FM} is imperative, as tasks can only (or, at least, more naturally) be described in terms of natural language.

Moreover, generalisability does not manifest in a vacuum. 
Rather, it emerges as the model exploits the prior knowledge acquired during training.
A quintessential case of in-context learning is asking an audio \ac{FM} to caption an input audio stream.
For example, we might ask it to ``Describe all the objects and interactions that are audible in the input audio and take place within 10 meters of the microphone''.
That this capability is emergent, boils down to three things:
a) It has (obviously) never `heard' the actual input audio before, nor encountered the particular combination of `objects' and `interactions' that is included in it;
b) It has never been trained to perform the task of `describe all objects and their interactions';
c) It has potentially never encountered any, or at least some, of the included `objects' or `interactions' in audible form during its pretraining (\cf \cref{ssec:pretraining}); it has only encountered it in its language pretraining.
Yet, despite the fact that a model has never been trained to perform the requested task nor has ever heard any of the requested sounds before, it may be able to answer the query satisfactorily.

Where does this magic answer come from?
On a first level, the model needs to linguistically process the input query; this is a capability it inherits from its language module.
It also needs to be able to decompose individual sounds; this is done by the audio module.
Crucially, it needs to connect audio to some linguistic representation, such that it can combine that representation with the input query, understand what is being asked from it, and answer correctly.
This will allow it to leverage prior knowledge about how an unencountered audio source actually sounds -- even in the extreme case where its audio model has not seen it in its pretraining data.

For the purposes of this section, we distinguish between \acp{FM} with `native' support for audio and listening, and others where this ability is `unlocked' with additional training.
The first class of models is trained with multimodality in mind -- the models consume multimodal inputs already during pretraining.
In the second class of models, support for audio is an \emph{add-on} -- it is added to \acp{LLM} after training, usually by `connecting' an audio encoder to its input.
As mentioned, the reasons for this distinction are a) historical, as \acp{FM} first found success for language tasks and only then were coopted for audio, and b) computational, as most off-the-shelf pretrained \acp{FM} are in-essence `just' \acp{LLM}, so in order to give them the ability to hear, one must connect additional modules to them.

\subsubsection{Connecting audio and language models}
In the simplest terms, an interface between an audio and a language model is a differentiable module that maps the internal representations of the audio model to the input or intermediate (or both) representation space of the language model, i.\,e., putting audio into (sub-)words (also called ``tokens'' in contemporary lingo).
This mapping allows the language model to treat the input audio stream as if it was actual language.
It can thus connect it to the preceeding or proceeding words (denoting the prompt or query), map it to its internal representation space (denoting its internalised knowledge), and finally use it to produce the corresponding output (e.\,g., the caption of an input audio stream).
This connecting, differentiable module must be trained using standard \ac{DL} procedures.
It requires a finetuning dataset (made of audio-query-answer triplets), with the module trained to perform the mapping from audio to tokens.

The technical details of how this connection is realised \change{differs}{differ} across architectures.
Some choose \remove{for} simple mapping mechanisms (like a single linear layer or a small \ac{MLP}), whereas other employ larger sub-networks (e.\,g., using transformers).
In terms of \emph{where} the auditory information is injected, some works choose to interleave the mapped audio representations with the input linguistic tokens, while others inject it (also) into the intermediate layers.

According to a recent taxonomy~\citep{Wadekar24-TEO}, there are two main choices for this fusion: 1) internal or \emph{deep fusion} and 2) input or \emph{early fusion}.
A conceptual diagram for both methods is depicted in \cref{fig:foundation}.
Both start from an existing \ac{LLM} and introduce new, multimodal capabilities into it by injecting multimodal information.
Deep fusion methods inject multimodal information within one or more layers of an existing \ac{LLM}, whereas early fusion does so only at the input level.

This taxonomy can be further broken down to four types of fusion depending on \emph{where} the fusion happens (at the input stage vs the intermediate layers) and \emph{how}:
\textbf{Type A}: This corresponds to a deep fusion using the nowadays `standard' cross-attention mechanism for modality integration, i.\,e., the scaled dot-product attention introduced in \citet{Vaswani2017-AIA}.
\textbf{Type B}: This is a variation of the above, where alternative fusion methods are considered. These are still attention-based, but introduce additional learnable components, such as gating functions or positional encodings. 
\textbf{Type C}: This first type of early fusion introduces multimodal information before the \emph{tokenisation} stage. This requires the modality-specific encoders to learn a mapping to the tokenised space already learnt by the \ac{LLM}. This allows to use the pretrained \ac{LLM} without finetuning.
\textbf{Type D}: In contrast to the above, the different modalities are tokenised separately for each type of fusion. While this requires finetuning the \ac{LLM}, it has the added benefit of being able to produce multimodal tokens, allowing for multimodal generation capabilities (see \cref{ssec:generation}). 
For an extended analysis of the advantages and disadvantages of these methods, see \citet{Wadekar24-TEO}.
For our purposes, we use only the higher-level taxonomy, namely, we categorise models as following an early or a deep fusion.

Finally, there exists, of course, a plethora of different strategies for finetuning:
While the mapping module \emph{must} be trained by necessity, the audio and language modules can be frozen or jointly finetuned; moreover, this joint finetuning may happen in stages, with the mapping module trained first and the other two remaining frozen, followed by a subsequent unfreezing of one or both of them.
Simpler tasks may be preferred in the beginning (e.\,g.,  \ac{ASC}), followed by progressively harder, more complex tasks, like \ac{AC}.

Generally, the optimal way for connecting audio and language models is still a very active area of open research.
Rather than aiming to document all available methods, we choose to describe the basic principles that underpin them (but see \citet{Latif23-SOL} for a recent overview).
Despite marginal differences, all methods ultimately follow the same recipe -- to align learnt audio and linguistic representations, thus unlocking prompting and instruction-following capabilities for audio \acp{FM}.

\subsubsection{Native multimodal support}
\label{ssec:native}
Besides audio \acp{FM}, which introduce an additional module to map audio inputs to tokens that are then passed to a pretrained \ac{LLM}, there are also some -- increasingly more -- models with native support for audio.
These models operate under the same fundamental principles: they accept multimodal prompts as input and generate their (multimodal) outputs autoregressively.
The main difference lies in how those models are pretrained.
Instead of training the audio and linguistic modules separately, they are \emph{subsumed} into a single, monolithic architecture that views both as tokens.

\subsubsection{\add{Multichannel support}}
\add{
As we show in \mbox{\cref{ssec:existing}}, most existing {\acp{FM}} only support single-channel inputs.
BAT~\mbox{\citep{Zheng24-BLT}} is the only notable exception.
Processing multichannel audio and extracting spatial information from it requires adaptation of the audio encoder.
Information from the multiple channels must be integrated at some point in the architecture.
BAT does so at the input stage by incorporating the phase difference between two channels.
In principle, it is also possible to fuse spatial information in any arbitrary intermediate point within a network, similar to how multimodal information is integrated.
However, as noted, a key challenge in spatial process is the relative scarcity of data, which becomes even more pertinent in the case of data-hungry {\acp{FM}}.
We anticipate the use of acoustics simulators, such as \emph{pyroomacoustics}~\mbox{\citep{Scheibler18-PAP}} or \emph{gpuRIR}~\mbox{\citep{Diaz21-GAP}}, to play an important role in developing {\acp{FM}} with spatial audio capabilities.
}

\subsubsection{\add{In-context learning: }From finetuning to prompting}

After discussing how audio capabilities are added to \acp{LLM} (or built into natively multimodal \acp{FM}), we continue with an outline of how the resulting \emph{multimodal}\add{, \emph{Tier-2}} foundation models can be used to perform audio tasks.
Here, the dominant paradigm has changed from \emph{finetuning} to \emph{prompting} -- the adaptation of the input sequence to match the target task.
In this subsection, we give an overview of prompting, both its most common, linguistic form, and auditory prompting, and how it helps achieve the goal of audio understanding.

\textbf{Language prompting:} Prompting emerged primarily in the domain of \acp{LLM}.
Designing prompts for downstream tasks guides the \acp{LLM} to achieve a particular objective by conditioning the probability of the next token prediction on additional information (the `prompt').
This method, known as \emph{in-context learning}, has marked a pivotal transition in how models are treated.
In the context of \acp{LLM}, prompts are phrases (or even several sentences) which serve to a) provide additional information (`context') to the \acp{LLM} as it makes its predictions, for instance, by providing examples of how each input should be treated, and b) guide it to produce a suitable output.
Some prompt examples are shown in \cref{tab:prompts}.


\begin{table*}[t]
    \centering
    \caption{
    Audio analysis prompt templates with examples.
    \add{These are examples of prompts that can be used in conjunction with \emph{Tier-2} audio {\acp{FM}} to specify the task that the model should solve.}
    \add{[AUDIO] specifies the raw audio samples that are passed to the model along with the input prompt.}
    }
    \label{tab:prompts}
    \begin{tabular}{cl}
    \toprule
    \textbf{Type} & \textbf{Prompt} \\
    \midrule
    Free-form & Describe the audio content of the following audio clip: [AUDIO] \\
    \midrule
    Instruction & Describe the timbre and pitch of the following audio clip: [AUDIO] \\
    \midrule
    Few-shot  & The following are examples of audio clips and their descriptions: \\
              & [AUDIO$_\text{1}$] => ``The sound of rain falling'' \\
             & [AUDIO$_\text{2}$] => ``Someone typing on a keyboard'' \\
             & [AUDIO$_\text{3}$] => ``The continuous hum of cars passing on a busy street'' \\
              & Describe the sound of [AUDIO] => \\
    \bottomrule
    \end{tabular}
\end{table*}


\textbf{Audio-language prompting:}
Ultimately, audio tasks require the use of audio-linguistic prompts; one part of the prompt corresponds to the audio to be analysed, and the rest to the task that needs to be executed, as well as any additional context that the user wishes to codify (e.\,g., the style of response or available metadata about the audio).
As mentioned, the simplest way is to concatenate (or interlace) the tokens coming from the text and audio encoders before propagating them to the \ac{LLM} backbone, but this integration can also happen in the intermediate layers of the \ac{LLM} (see \cref{fig:foundation}).
In fact, this early fusion is the strategy that most audio \acp{FM} follow (see \cref{ssec:existing}), as it allows them to benefit both from the compositionality and ease-of-use that language offers and condition the output on acoustic properties using suitable audio prompts or references.

\subsubsection{Evaluating response appropriateness}

Beyond the evaluation of audio \acp{FM} on standardised benchmarks, which essentially follows the same procedure as traditional methods, their scalability and \emph{open-endedness} introduce additional challenges.
Specifically, benchmarks comprising labelled data can evaluate whether a response is \emph{correct} or not, but they cannot easily capture whether it is \emph{appropriate}.

This \emph{appropriateness} becomes an important topic once we transition to open-ended captioning (one of the newly-emergent tasks of audio analysis).
While a model can be trained to produce a rich and faithful caption, the question of whether its response should \emph{contain} all the sources, events, actions, and ontologies it is able to recognise remains open.
Ultimately, this decision rests with the designer of the application that employs an audio analysis component.
While the audition module of an autonomous driving agent should provide as much information about its environment as possible, a conversational assistant in a smartphone might, presumably, hide some of that information to make its response more easily digestible to a human user.

While the models themselves can technically be agnostic to which parts of their output are being used by downstream applications, the issue of evaluating their response according to the criteria of appropriateness still remains an open problem.
For \acp{LLM}, this problem has been circumvented by collected large corpora of annotated human-machine conversations (which can also be used to finetune the models).
We expect similar resources to emerge for the evaluation of audio \acp{FM} in the near future, similar to \citet{Zheng24-BLT}, \citet{Gong24-LTA} and others who curate large, text-audio pairs for the training and evaluation of their models.

\subsubsection{Improving appropriateness}
\label{ssec:appropriate}
\change{Foundation models}{\emph{Tier-2} {\acp{FM}}} are considered to be task-agnostic -- at least in their original state before finetuning.
At this stage, the models have been exposed to massive amounts of unlabelled data and trained to model the underlying data distribution in self-supervised fashion.
While this already results in powerful models that can produce generic outputs, there is almost always a need to further finetune them such that they respond in particular ways.
For audio, this means adapting the model to tackle the tasks that are outlined in \cref{sec:tasks}.
The two most widely-used methods to achieve this adaptation are \emph{instruction finetuning} and \emph{retrieval-augmented generation}.
We outline both in the subsections that follow and outline how these two steps can be circumvented by \emph{merging} different components.

\begin{figure*}[t]
    \centering
    \includegraphics[width=\textwidth]{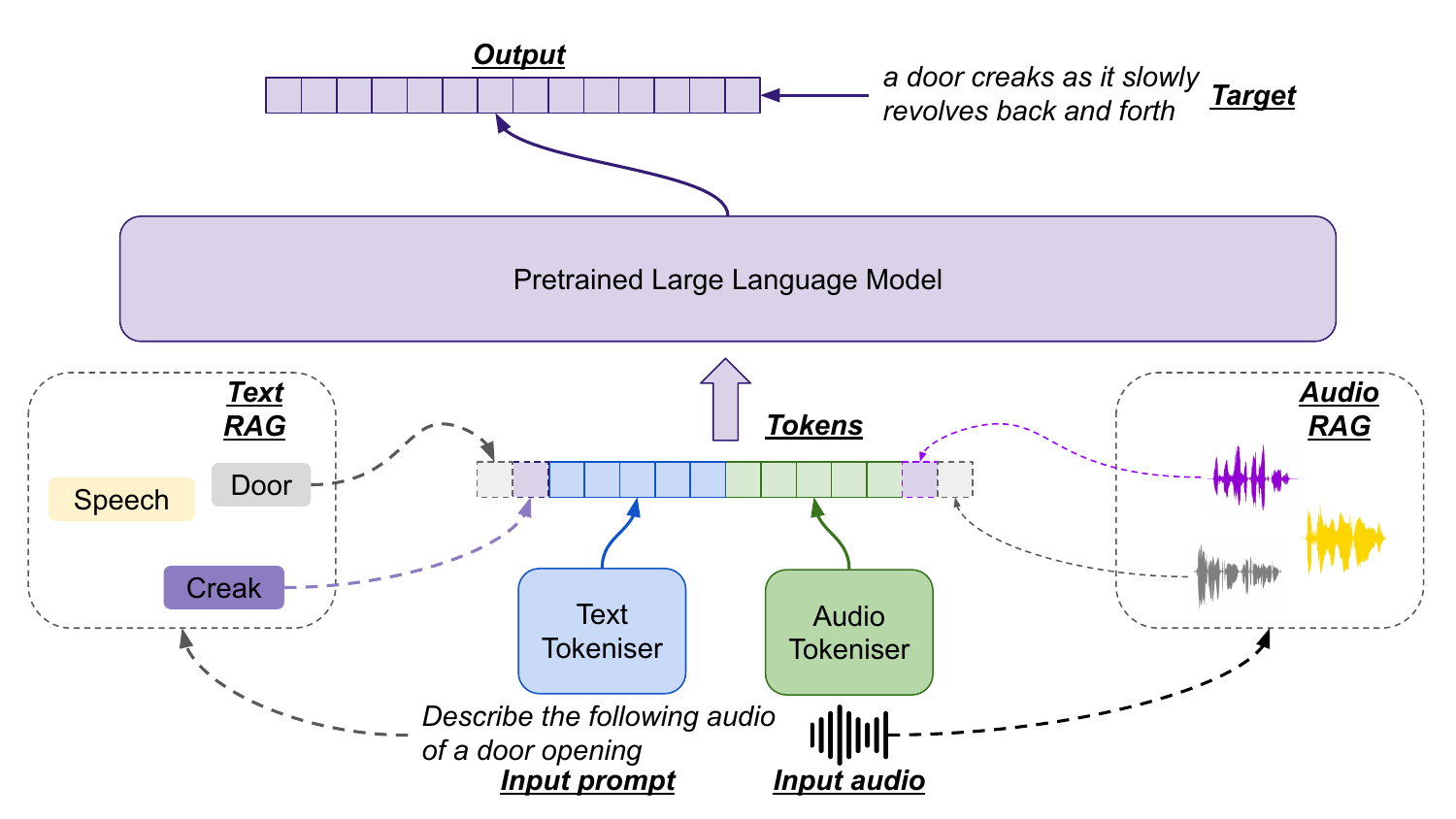}
    \caption{
    Overview of retrieval-augmented generation (RAG) for the case of early fusion.
    The initial audio and text prompts are used as queries for external databases (separately for audio and text).
    Retrieved queries are encoded/tokenised and concatenated to the initial prompt, which is then fed to the pretrained model.
    }
    \label{fig:rag}
\end{figure*}

\textbf{Instruction finetuning:}
\emph{Instructions} are targeted prompts which specify how exactly the model output should look like~\citep{Chung24-SIL}.
During instruction finetuning, the model is trained with (instruction, output) pairs using standard supervised learning.
Generating those pairs is done either by painstakingly collecting human annotations or, increasingly, by using existing (foundation) models to automatically generate those responses~\citep{Wang23-SEL}.
In the context of instruction finetuning, we begin with a labelled dataset $\mathcal{D} = \{(\mathbf{x}, \mathbf{y}), \text{ instruction }\mathbf{x}, \text{ output } \mathbf{y}\}$, where $\mathbf{x} = (\pmb{x}_1, \pmb{x}_2, \ldots, \pmb{x}_k)$ represents the instruction (input tokens) and $\mathbf{y}$ represents the output (label), which is also represented with tokens.
Similar to \change{\mbox{\cref{eq:loss_function}}}{autoregressive modeling}, the objective is to maximise the corresponding likelihood\change{ as defined by \mbox{\cref{eq:instruction_loss}}:}{.}
\remove{with prior research indicating that a linear combination of \mbox{\cref{eq:loss_function}} and \mbox{\cref{eq:instruction_loss}} yields improved performance~\mbox{\citep{Radford18-ILU}}.}

Crucially, instruction tokens are interlaced with other tokens, with the distinction being that instruction tokens remain consistent across all training and inference instances.
This distinction is more clear in the case of \acp{LLM}.
There, one can use the instruction tokens to specify the exact form that the output of the \ac{LLM} should take.
For instance, the phrase ``Answer the question with `FOO' or `BAR'.'' can be prepended to a set of training questions, with each question itself becoming part of the prompt.
The \ac{LLM} is then trained to produce FOO/BAR answers (given labelled data).
This `unlocks' the ability to answer such questions during inference (i.\,e., questions not seen during training).
Importantly, the \ac{LLM} will only answer the question in the exact format specified if -- and only if -- the instruction is prepended to the form during inference.
Thus, $\mathbf{x}$ can be broken down to $\mathbf{x_i}$ (the instruction tokens) and $\mathbf{x_p}$ (the prompt tokens).

As we see in \cref{ssec:existing}, this distinction is not so relevant for the case of audio \acp{FM}.
This is because audio \acp{FM} are -- to a large extent -- `simply' \acp{LLM} that have been instruction finetuned to answer audio queries.
They are trained to respond to particular tasks such as ``Describe the audio content of the following clip:'' (with the audio tokens concatenated to this textual instruction).
This means that the difference between \emph{prompts} and \emph{instructions} is not so clear for audio (or, generally, multimodal) \acp{FM}.


Finally, due to the sheer number of parameters, adapting the weights of \acp{FM} can be a computationally expensive process -- oftentimes prohibitive for smaller research groups with limited resources.
For this reason, the community has spent considerable effort on advancing \ac{PEFT} methods. 
Typical examples include \ac{LoRA}~\citep{Hu22-LOR} and its quantised version, QLoRA~\citep{Dettmers23-QEF}, which parametrise weights into low-dimensional subspaces and proceed to update those subspaces instead of the original weights.
This streamlines the process of instruction finetuning enabling the training of \acp{FM} that can respond to specific prompts in the desired way -- with the crucial requirement of a good quality dataset that contains a mapping between prompts and responses.



\textbf{Retrieval-augmented generation:}
\Ac{RAG}~\citep{Lewis20-RGF} has emerged as an indispensable tool within the landscape of \acp{FM}.
Defined as the integration of external knowledge into the generative process, \ac{RAG} addresses the challenge that \acp{FM} face with handling domain-specific or highly specialised queries which were \emph{absent} from their training data.
These challenges often lead to inaccuracies (or ``hallucinations'') in the generated outputs.

In general, \ac{RAG} leverages external \emph{assets} (i.\,e., data that are relevant for the target task).
These assets are available in the form of a queryable database that returns the most relevant responses to particular queries.
The query is based on the input prompt -- either all of it, or part of it.
The response to the query is then incorporated into the prompt that is given to the \ac{FM}.
While -- in principle -- any type of database would serve (e.\,g., a(n) (No)SQL database), there are two important specifications: a) the database must be queryable through the same data types that the \acp{FM} can be prompted with (as it is the prompt itself that becomes the query), and b) the response of the database must also be of the same data type as the prompt (as it will be incorporated into the prompt and, thus, \emph{become} the prompt).
In simple words, the databases that are used for \acp{LLM} must support text queries and return text responses, whereas the databases used for audio \acp{FM} must support audio (and language) queries and return audio (and language).

An illustrative example is shown in \cref{fig:rag} (for the case of early fusion; see \cref{ssec:interfaces}).
A trained audio \ac{FM} is prompted to describe the sound of a door closing, and given a clip of that audio event.
A \ac{RAG} pipeline then uses the textual part of the prompt (``Describe the following audio of a door opening'') to query an external text database (e.\,g., a database of common audio tags).
The query is compared to each asset in that database, and relevant assets are returned.
They are tokenised in the same format as the original prompt and concatenated to it.
Similarly, the audio part of the prompt (the audio clip itself; potentially tokenised already) is used to query a database of audio assets, the most relevant of which are returned, tokenised, and concatenated to the prompt.
The resulting prompt -- which comprises the original plus the tokenised representations of the assets returned by \ac{RAG} -- is then given to the \ac{FM} so that it can produce its output, which is now informed by extra information.

\textbf{Language \ac{RAG}:} In the case of language, \ac{RAG} bridges the gap between the requested information and the knowledge that an \ac{FM} has incorporated during training by querying an external database~\citep{Lewis20-RGF}.
Given the success of \acp{LLM} in understanding text, contemporary \ac{RAG} approaches often rely on them to perform the queries, as seen in the examples of LlamaIndex~\footnote{\url{https://www.llamaindex.ai/}} and LangChain~\footnote{\url{https://www.langchain.com/}}.
There, an \ac{LLM} (oftentimes the same which is being augmented with \ac{RAG}) is used to extract embeddings (i.\,e., some well-performing, intermediate representation of that \ac{LLM}) from all external textual assets.
During inference, the prompt is first encoded using the same \ac{LLM}, and its embeddings are compared with the embeddings already extracted offline.
The comparison can be simple (cosine similarity) or more elaborate.
The assets are ranked, and some policy is used to retrieve the most relevant ones (e.\,g., those with the top-5 highest similarity).


\textbf{Audio \ac{RAG}:} 
\ac{RAG} can also be incorporated to handle audio queries, leading to a form of multimodal \ac{RAG}.
Specifically, rather than (or in conjunction to) querying external text assets, the same concept can be used to query \emph{audio assets} as well.
In the simplest case, the query can be used to retrieve textual content related to the input audio (e.\,g., captions).
For instance, \citet{Ghosh24-RRA} retrieve audio related to the input and subsequently append to the prompt the captions associated with these retrieved audio snippets.
However, we can envision the introduction of the audio snippets \emph{themselves} (or, to be precise, their representation) as part of the prompt (along with textual information related to it).
This can help the \ac{FM} directly associate parts of its audio query with the related parts of a number of other audios, thus further enhancing the appropriateness of the response.

\begin{figure}[t]
    \centering
    \includegraphics[width=\columnwidth]{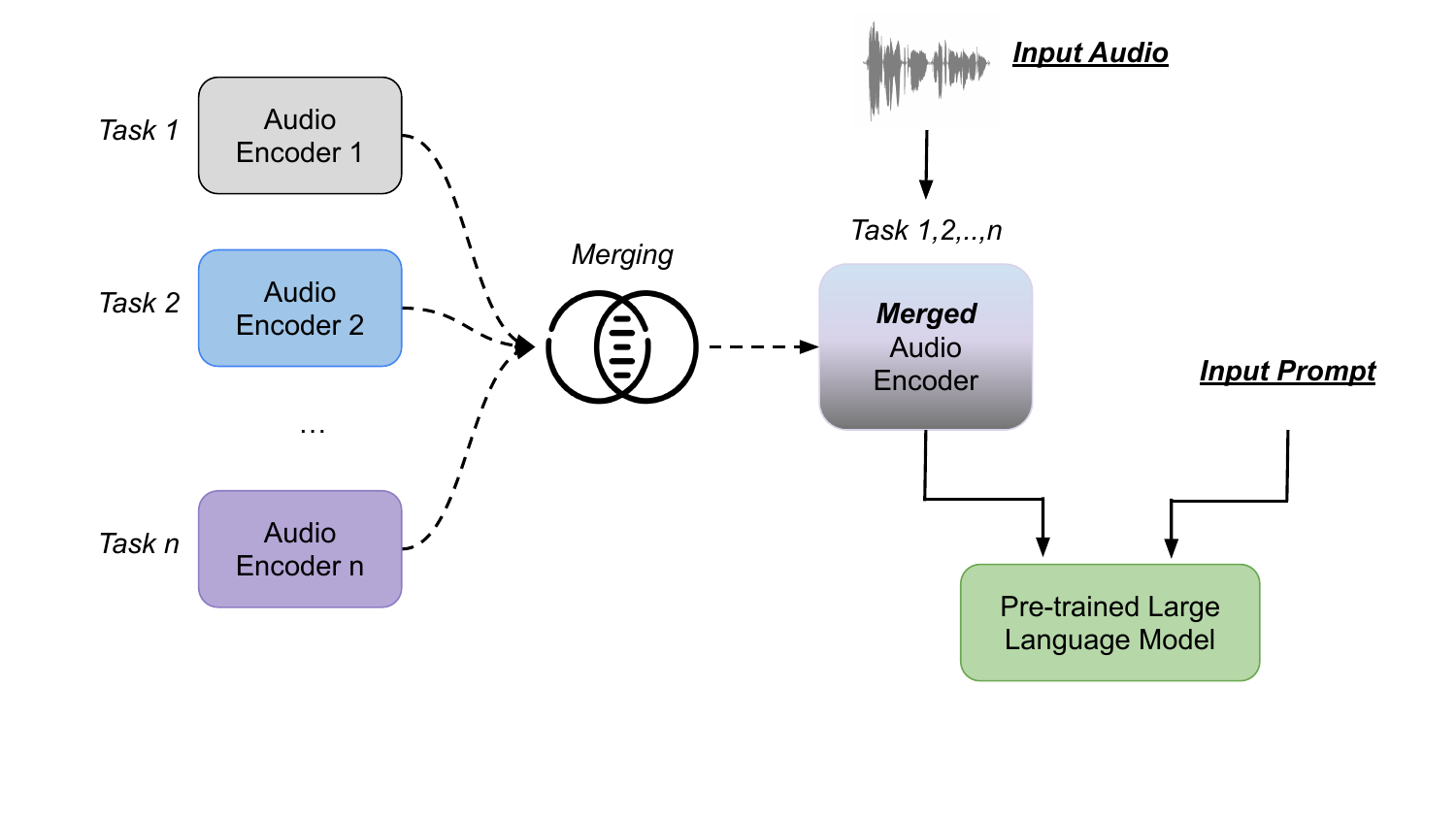}
    \caption{
    Merging process of audio language models.
    The process of merging can be seen as an \emph{interpolation} in weight space across one or more pretrained models.
    This combines the abilities of multiple models with little to no additional training.
    Merging task-specific trained encoders and attaching the merged one as the audio encoder in the audio language model is anticipated to provide improvements in the domain in the future.
    }
    \label{fig:mlm-merge}
\end{figure}

\textbf{Merging unimodal foundation models:}
Another way of unlocking new abilities in \acp{FM}, beyond instruction-finetuning, is that of \emph{merging}\add{~\mbox{\citep{Utans96-WEI, Entezari21-ROL, Goddard24-ARC}}}, shown in \cref{fig:mlm-merge}.
Merging \acp{FM} essentially involves the task of combining the weights of multiple models into a single, unified model.
There are various techniques for achieving this, each aiming to leverage the strengths of individual models (either sharing the same architecture or not) to create a more robust and versatile system.
This approach is particularly beneficial given the complexity and resource-intensity of training \acp{FM}.

The process of merging \acp{LLM} involves various techniques and has led to several state-of-the-art models~\citep{Goddard24-ARC}. One fundamental approach is \ac{LMC}~\citep{Entezari21-ROL}, which combines \emph{models with identical architectures and initialisations} through linear interpolation of weights. Linear averaging, as discussed by~\citet{Utans96-WEI} and expanded by~\citet{Wortsman22-MOD} in the ``Model Soups'' approach, forms the basis of many merging techniques by leveraging the similarities in weight space. 
For \emph{models with identical architectures but different initialisations}, the permutation symmetry of neural network checkpoints is crucial. Techniques like Git-Rebasin~\citep{Goddard24-ARC} align weights from independently trained models to achieve functionally equivalent configurations. This alignment facilitates effective merging despite differences in initial training paths.
Similarly, approaches such as \citet{Tatro20-OPT} reduce the interpolation barrier by assigning correspondences between neurons in the models.
Advanced methods also support \emph{merging models with different architectural configurations}. For instance, the \ac{CALM} method~\citep{Bansal24-LLM} employs cross-attention mechanisms to integrate representations from diverse models, leveraging their combined strengths, whereas FUSELLM~\citep{Wan24-KNO} aligns and fuses probabilistic distributions of source \acp{LLM} to enhance their generative capabilities; however, it requires additional pretraining.

Overall, by merging off-the-shelf pretrained models specialised in different tasks, one can develop a single, comprehensive model capable of effectively handling a diverse range of tasks.
This has been primarily explored for the case of \acp{LLM}, but is, in principle, also possible for audio \acp{FM} -- that is, as an independent step before connecting audio and language models.
While with the current state of the art, it is only possible to merge \acp{FM} from the same modality, we cover the merging of multimodal \acp{FM} as a future direction in \cref{ssec:merging:multimodal}.

\subsection{Existing audio foundation models}
\label{ssec:existing}

After outlining the inner workings of audio \acp{FM}, we present a list of models with \add{\emph{Tier-2}} audio listening capabilities in \cref{tab:overview}.
We note that this list is not meant to be exhaustive; indeed, with the pace that this field is developing, we expect it to become obsolete within a few weeks after publishing our work.
Instead, we aim to demonstrate the similarities and differences in the data, pretraining tasks, and architectures of existing modules, in order to gain an understanding of the design choices that have been considered so far.
In the subsections that follow, we give an overview of their most important components.

\afterpage{
\begin{landscape}
\begin{table}
\centering
\caption{
Overview of existing \emph{Tier-2} audio-language models.
Our overview includes the main differentiable modules of each architecture (\textcolor{blue}{\ding{100}} means the module is frozen, whereas \textcolor{orange}{\faFire} means it is finetuned), the data used for pretraining, the pretraining objective, the type of architecture (see \cref{sec:multimodal}), and the capabilities claimed by the original authors.
In this version of the table, we only include the data and capabilities relevant to our work here (i.\,e., that maps to the tasks we consider in \cref{sec:tasks}).
See \cref{app:existing} for the full-version of the table including all tasks and capabilities (e.\,g., speech and music).
}
\label{tab:overview}
\begin{threeparttable}
\resizebox{\textwidth}{!}{\begin{supertabular}
{p{0.2\textwidth}|p{0.3\textwidth}p{0.4\textwidth}p{0.1\textwidth}|p{0.4\textwidth}}
\toprule
\textbf{Model} & \textbf{Modules} & \textbf{Data Used} & \textbf{Fusion Type} & \textbf{Capabilities}\\
\midrule
\multicolumn{5}{c}{\textbf{Publicly Available}}\\
\midrule
Pengi~\citep{Deshmukh23-PAA} & Audio encoder\textcolor{orange}{\faFire}: CLAP, \newline Text encoder\textcolor{blue}{\ding{100}}: CLAP \newline Audio and Text Prefix Layers\textcolor{orange}{\faFire} \newline LLM\textcolor{blue}{\ding{100}}: GPT2-base & AudioSet, FSD50K, CochlScene, AudioCaps, Clotho (v2), ClothoAQA, WavText5K, SoundDescs, MACS, WavCaps, Freesound, FindSound$^*$ & Early & Audio captioning, Audio question answering, Sound event classification, Acoustic scene classification, Audio-language retrieval\\

\midrule
APT-LLM~\citep{Liang23-APT} & Audio encoder: Audio-MAE\textcolor{blue}{\ding{100}} \newline Audio aligner\textcolor{orange}{\faFire} \newline 
Acoustic adapters\textcolor{orange}{\faFire}: APT \newline LLM\textcolor{blue}{\ding{100}}: Vicuna-7b (v1.1)$^\dag$ & AudioSet, Wavcaps, AudioCaps, Clotho (v2), Clotho AQA, AudioSet-SL, NLAR & Early & Audio tagging, Sound event classification, Natural language-audio reasoning, Audio-visual question answering\\
\midrule
LTU~\citep{Gong24-LTA} & Audio encoder\textcolor{orange}{\faFire}: AST \newline LoRA adapters\textcolor{orange}{\faFire} \newline LLM\textcolor{blue}{\ding{100}}: LLaMA-7b & OpenAQA
& Early & Audio tagging, Audio captioning, Audio reasoning, Audio temporal analysis, Sound event classification
\\
\midrule
LTU-AS~\citep{Gong23-JAA} & Audio encoder\textcolor{blue}{\ding{100}}: Whisper \newline TLTR\textcolor{orange}{\faFire} \newline
Projection layer\textcolor{orange}{\faFire}\newline LoRA adapters\textcolor{orange}{\faFire} \newline LLM\textcolor{blue}{\ding{100}}: LLaMA-7b & Open-ASQA & Early & Audio captioning, Sound event classification, Audio question answering, Non-speech audio event recognition\\
\midrule
QWEN-Audio~\citep{Chu23-QAU} & Audio encoder\textcolor{orange}{\faFire}: Whisper-L-v2, \newline LLM\textcolor{blue}{\ding{100}}: Qwen-7b & N/A &  Early & Audio captioning, Sound event classification, Acoustic scene classification, Sound event detection, Audio question answering \\
\midrule
Salmonn~\citep{Tang24-STG} & Audio encoders\textcolor{blue}{\ding{100}}: Whisper, BEATs, 
\newline Q-Former\textcolor{orange}{\faFire} \newline 
LoRA adapters\textcolor{orange}{\faFire} \newline
LLM\textcolor{blue}{\ding{100}}: Vicuna & WavCaps, AudioCaps, Clotho & Early & Audio captioning, Audio question answering\\
\midrule
BAT~\citep{Zheng24-BLT} & Audio encoder\textcolor{blue}{\ding{100}}: Spatial-ASL, \newline Projection Layer\textcolor{orange}{\faFire} \newline LLM\textcolor{orange}{\faFire}: LLaMA2 & SpatialSoundQA & 
Early & Sound event classification, Sound event detection and spatial localisation, Audio question answering\\
\midrule
Audio Flamingo~\citep{Kong24-AFA} & Audio encoder\textcolor{blue}{\ding{100}}: CLAP \newline Projection Layer\textcolor{orange}{\faFire} \newline LLM\textcolor{orange}{\faFire}: opt-iml-max-1.3b& WavCaps, Macs, SoundDescs, Clotho-v2, WavText5K, LAION-630k, Clotho-AQA, Open-AQA, AudioSet, FSD50k, CochlScene, NonSpeech7K, Chime-Home, Sonyc-UST & Deep & Audio captioning, Sound event classification, Audio question answering \\
\midrule
UniAudio 1.5~\citep{Yang23-UAA} & Audio encoder\textcolor{orange}{\faFire}: LLM-Codec \newline LLM\textcolor{blue}{\ding{100}}: LLaMA2-7b & AudioCaps & Early & Sound event classification\\
\midrule

\multicolumn{5}{c}{\textbf{Restricted Access}}\\
\midrule
Gemini 1.5 Pro~\citep{Reid24-GUM} & N/A & N/A & N/A & Audio understanding, Audio captioning\\
\midrule
GPT-4o\footnote{\url{https://openai.com/index/hello-gpt-4o/}} & N/A & N/A & N/A & Audio understanding, Audio captioning\\
\midrule
LauraGPT~\citep{Chen23-LLA} & Audio encoder\textcolor{orange}{\faFire}: Encodec-based \newline LLM\textcolor{orange}{\faFire}: Qwen & Clotho, AudioCaps, WavCaps, FSD-50K & Early & Audio captioning\\
\bottomrule
\end{supertabular}}
\begin{tablenotes}
\item[*] It is not clear what data was exactly used from FindSounds.
\end{tablenotes}
\end{threeparttable}
\end{table}
\end{landscape}
}
\afterpage{
\begin{landscape}[t]
\centering

\tablefirsthead{%
\toprule
\textbf{Dataset} & \textbf{Comments} \\ \midrule \\
}
\tablehead{%
\toprule
\textbf{Dataset} & \textbf{Comments} \\
}
\tabletail{%
\bottomrule
\multicolumn{2}{c}{\small continued on next page} \\
\bottomrule
}
\tablelasttail{%
\bottomrule
}
\tablecaption{
    Overview of datasets used to train existing audio-language \acp{FM} along with some qualitative observations.
    Wherever relevant, we also denote the source of each dataset, thus giving a sense of their lineage.
    Note that some interpretations here are necessarily subjective (e.\,g., there is no threshold for small- vs large-scale, but we nevertheless wanted to give a measure of each dataset's size).
}
\label{tab:datasets}

\begin{supertabular}{p{.2\textwidth}|p{.8\textwidth}}
        AudioSet~\citep{Gemmeke17-ASA} \newline (Source: YouTube) & AudioSet is a large (ca. 2M instances) audio(visual) dataset collected from YouTube. Original user tags have been verified by human raters according to a hierarchical ontology of 527 categories. Each instance corresponds to a 10-second clip that features multiple labels (on average 2.7), thus showing a low degree of polyphony. Note that the original dataset includes the IDs of original videos and not the videos themselves; given that content is constantly removed from the platform, the AudioSet dataset has been consistently shrinking in size. Moreover, different groups may have slightly different versions of the dataset depending on the date of download. The dataset also partially suffers from missing labels~\citep{Fonseca20-AML}.\\ 
        \midrule
        AudioSet-SL~\citep{Hershey21-BEN} \newline (Source: AudioSet) & This is a smaller subset (67k instances) of the AudioSet dataset which has been further annotated on a frame-by-frame level for all events present in a clip. \\ 
        \midrule
        Freesound~\citep{Fonseca17-FDA} & Freesound is an online repository for sounds, where users can upload audio recordings with multiple tags and short descriptions. It currently features over 600k short clips. Recordings tend to be focused on a single source (with no or minimal presence of other sources).\\
        \midrule
        FSD50K~\citep{Fonseca21-FAO} \newline (Source: Freesound) & FSD50K features a small (ca.\ 51k instances) curation of clips from Freesound that have been weakly annotated by human raters according to a subset of the AudioSet ontology (200 classes).\\ 
        \midrule
        CochlScene~\citep{Jeong22-CAO} & CochlScene is a crowdsourced dataset collected from smartphone users in Korea. Users were asked to submit data conforming to a small set of labels using their mobile phones. They were further asked to verify data collected from other users.
        \\ 
        \midrule
        AudioCaps~\citep{Kim19-AGC} \newline (Source: AudioSet) & AudioCaps comprises a smaller (ca.\  46k) subset of AudioSet with human-generated captions. Annotators were given word hints from the original AudioSet classes and had access to the video stream when captioning (instructed to use it only when unsure about the audio content)\\ 
        \midrule
        Clotho~\citep{Drossos20-CAA} \newline (Source: Freesound) & Clotho contains captions for ca.\ 5k instances from Freesound. Raters were given no hints during the annotation process.\\ 
        \midrule
        ClothoAQA~\citep{Lipping22-CAC} \newline (Source: Clotho) & ClothoAQA is a subset (ca.\ 2k instances) of Clotho that features question-answer pairs. It only contains  questions with yes/no and single-word answers. The questions were formulated by human annotators in a three-stage process; raters of later stages had access to the questions from the previous stages and were instructed to formulate different ones. Each question was answered by a different annotator than the one who phrased it.
        \\ 
        \midrule
        WavText5K~\cite{Deshmukh22-ARW} \newline (Source: BigSoundBank and SoundBible) & WavText5K is a small-scale dataset featuring sound clips sourced from two online repositories that have permissing licensing. The clips were annotated by the original uploaders. The repositories are marketed for ``sound effects'' specialists, thus presumably feature `prototypical' sounds collected with high-quality microphones and little interference or polyphony.\\ 
        \midrule
        SoundDescs~\citep{Koepke22-ARW} \newline (Source: BBC Sound Effects) & SoundDescs is similar to WavText5K, but collected from a different source with a non-commercial license.\\ 
        \midrule
        MACS~\citep{Martin21-WIT} \newline (Source: TAU Urban Scenes Dataset) & MACS is a small subset of an audio scene classication with human-annotated captions. It has a higher degree of polyphony than other caption datasets given the original emphasis on recording natural soundscapes. Only three underlying soundscape classes were used. The original audio was collected using a small set of recording devices and recordings were situated in a few European cities -- though from numerous locations within them.\\ 
        \midrule
        WavCaps~\citep{Mei24-WAC} \newline (Source: Freesound, AudioSet-SL, SoundBible, BBC Sound Effects) & WavCaps is a large-scale (ca.\ 400k) automatically-captioned dataset. The captions were generated using ChatGPT (a version based on GPT-3) to reformulate the original tags or descriptions in the dataset's sources.        \\ 
        \midrule
        NLAR~\citep{Liang23-APT} \newline (Source: ClothoAQA) & After prefiltering the original annotations of ClothoAQA to remove erroneous annotations, the authors used ChatGPT to summarise the acoustic characteristics of each caption. Subsequently, they utilised ChatGPT to create questions pertaining to \emph{pairs} of original audios and their preprocessed captions.\\ 
        \midrule
        OpenAQA~\citep{Gong24-LTA} \newline (Source: AudioSet, AudioSet-SL, VGGSound, AudioCaps, Freesound, Clotho, Sound Bible) & 
        OpenAQA is a large, open-ended question-answer dataset.
        The dataset has been sourced from multiple domains, and the captions are automatically generated from the original tags/captions/descrptions using ChatGPT.
        \\ 
        \midrule
        SpatialSoundQA~\citep{Zheng24-BLT} \newline (Source: AudioSet) & 
        Audio data originally taken from AudioSet was rendered using the SoundSpaces 2.0 simulator~\citep{Chen22-SAS} with \acp{RIR} coming from 90 scanned buildings. Questions were created using templates and subsequently rephrased with GPT4.\\ 
        \midrule
        LAION-630k~\citep{Wu23-LCL} \newline (Source: Freesound, BBC Sound Effects, Free To Use Sounds, Sonniss Game Effects, We Sound Effects, Paramount Motion Sound Effects, Audiostock, Epidemic sound) &  This dataset is derived from its sources by aggregating sound clips and their accompanying -- uploader-defined -- descriptions.
        \\ 
        \midrule
        NonSpeech7K~\citep{Rashid23-NDC} \newline (Source: Freesound, YouTube, Aigei)  & NonSpeech7k is a curation of human non-speech sounds (e.\,g., coughing) that has been manually verified and strongly annotated on the frame level.
        \\ 
        \midrule
        Chime-Home~\citep{Foster15-CAD} & Chime-Home is a small dataset collected with a single device in a single environment. The recordings have been split in non-overlapping 4-second chunks, which are subsequently annotated by human raters for multiple sound events.\\ 
        \midrule
        Sonyc-UST~\citep{Cartwright19-SUS} & Sonyc-UST is a small-scale dataset of human-annotated tags. The data was recorded from multiple locations in New York City (USA), always using the same recording device.\\
\end{supertabular}
\end{landscape}
}

\subsubsection{Datasets}
\label{ssec:datasets}

It is evident from \cref{tab:overview} that different models often used different pretraining data -- in fact, we have not found a single \change{instances}{instance} of complete overlap in the training data used by two pairs of works.
This creates evident issues for comparability, which we return to later.

Given that statistical models will necessarily inherit the biases and limitations of the data they were trained on, it is important to highlight those limitations (and, also, the strengths).
We provide such a quick (partly subjective) overview in \cref{tab:datasets}.
Summarising the strengths and limitations, as well as the underlying sources of data in a succinct way, the following observations become evident:
\begin{itemize}
    \item Multiple datasets share the same underlying source, most notably, AudioSet and Freesound.
    While this is not a limitation, per se, it nevertheless means that all datasets derived from those two sources will inherit their biases.
    For both of these, a major source of concern is their lack of polyphony (e.\,g., AudioSet only has 2.7 labels per segment on average~\citep{Gemmeke17-ASA}).
    Coming from public domains where users upload their content of interest, they usually feature audio snippets where a particular source is very prominently in the foreground, with little or no interference from other, unrelated sources.
    While this is useful for training \ac{SED} models that recognise isolated events, it does not lead to the holistic and comprehensive understanding of a soundscape that is usually envisioned by the creators of the models we reviewed.
    \item Several data sources contain labels that have not been verified by experts, or at least contracted annotators.
    Instead, they come with user tags that were specified by the original uploaders to a public platform (e.\,g., YouTube or Freesound).
    While this reduces the annotation effort, it is nevertheless resulting in high label uncertainty. 
    \item As \acp{FM} that aim to tackle captioning require larger amounts of data than traditional models, and such data does not exist yet and requires considerable annotation efforts, several groups have taken to using pretrained \acp{LLM} (primarily ChatGPT-3.5-turbo) to create \ac{AI}-generated captions.
    While this can substantially increase the amount of labelled data, it is a considerable source of bias, as captions will follow the linguistic style of a particular \acp{LLM}.
    Moreover, this process is also heavily affected by the fact that some of the datasets do not come with human-annotated (or at least verified) labels, but rather with user-defined tags.
    This means that the captions will inherit the same errors that are present in those tags.
    \item In contrast to those large-scale datasets, there are also several small-scale ones, usually collected by academic groups in a constrained setup, primarily using a fixed recording device and collected over a few recording session across a limited number of locations.
    While these datasets come with a higher degree of confidence in their labels, they nevertheless suffer from their small scale.
    In particular, they only cover a very restricted set of recording conditions, rendering their generalisation across devices and locations questionable.
\end{itemize}

We note again the partially subjective nature of the preceding remarks; while it is beyond the scope of this work to thoroughly critique and compare all available datasets used for the training of audio \acp{FM}, it is nevertheless our intention to highlight that such shortcomings necessarily exist.
As a first step, we believe it would be important to thoroughly document those in the model cards that accompany the release of \acp{FM}~\citep{Mitchell19-MCF}.

\subsubsection{Audio \& language modules}
\cref{tab:overview} also makes it evident that there is large variability in the choice of audio and language components, with rare overlap across models.
One exception is Pengi~\citep{Deshmukh23-PAA} and Audio Flamingo~\citep{Kong24-AFA}, which use the same audio encoder; however, they use a different text decoder.
This is expected, as, in these early days of \acp{FM}, groups will typically co-opt the latest and best-performing components that are available to them.
On top of that, the relatively limited computational resources of most university groups make thorough `grid search' studies nearly impossible -- therefore, in their quest to achieve state-of-the-art results, newer works adopt the newest models that are available when they begin their experimentation.
This is why (we postulate) Pengi~\citep{Deshmukh23-PAA} used GPT2-base, but latter models used LLaMA2-7b~\citep{Touvron23-LOA}; we expect newer models to rely on LLaMA3, or even newer models in the future.

A critical requirement for audio encoders and tokenisers is the need to efficiently compress the information in the audio.
This is needed because \acp{LLM} struggle with longer content~\citep{Xiao24-ESL}.
Recent efforts have been aimed at creating more token-efficient codecs, such as EnCodec~\citep{Defossez23-HFN}, which can tokenise the input audio with very low rates (e.\,g., 50 tokens/s).
Finding such efficient representations is another current area of open research.

\subsubsection{\change{Pretraining}{Training} tasks}

The choice of pretraining task is another major design choice.
All models recast their target as a \emph{text generation} task by using templating to map classes or tags to a text string (e.\,g., ``This audio has the following tags: [TAG1], [TAG2]'' or ``The following audio belongs to [CATEGORY] class''.
Following this transformation, models are primarily trained on autoregressive text generation.
Some models, however, incorporate additional tasks, such as audio generation (in the case of UniAudio~\citep{Yang23-UAA} 1.5, where they use a generative-adversarial objective), transcription, and audio-text contrastive learning (e.\,g., APT-LLM~\citep{Liang23-APT} is trained to distinguish whether a pair of an audio recording and text are from the same source).
While introducing additional tasks is interesting, there is little understanding of how this helps, given the comparability issues described below.

\subsubsection{Capabilities}
Each released model comes with an assortment of capabilities claimed by its creators.
Typically, these capabilities are claimed based on benchmark performance on some dataset.
We note that these claims are rarely contextualised according to the strengths and limitations of each dataset -- which is another contribution of our work.
For example, some works claim ``Audio question answering'' capabilities while only evaluating on Cloth-AQA, which contains only a yes/no dataset.
This capability is more limited to open-ended audio-question answering as evaluated by OpenAQA~\citep{Gong24-LTA}.
While the issue of evaluation remains critical, it is evident from \cref{tab:overview} that audio \acp{FM} already exist that can handle most (but not yet all) of the audio analysis tasks we laid out in \cref{sec:tasks}.

\subsubsection{Comparing existing models}
\label{sssec:comparing}
The above subsections portray a distressing situation for comparing the capabilities of existing audio \acp{FM}.
As there is small overlap in the pretraining data and tasks, as well as the upstream audio and language models, it is hard to judge progress.
Crucially, the choice of training and testing datasets makes it impossible to compare some pairs of models.
For instance, Qwen-Audio~\citep{Chu23-QAU} is evaluated on CochlSchene and ClothoAQA, which were used in the training of other models, like Pengi~\citep{Deshmukh23-PAA}.
This is something we expect the community to solve in the next few years, for example, by making a concerted effort towards reproducible benchmarking.
We discuss this further in the next section.

\subsection{Multimodal foundation models}
\label{sec:multimodal}

Even though we have primarily focused on audio-language \change{foundation models}{{\acp{FM}}} in this article,
the principles we have covered can be extended to multimodal \acp{FM} that encapsulate additional modalities.
The standard one is \emph{vision}, which concerns the introduction of image or video processing capabilities to an \ac{FM} (or both), but any other number of sensors or actuators can be added as well.
The basic principle is the same -- these capabilities are either \emph{built-in} by training multimodally from scratch, or are added later by attaching an additional encoder that maps data from a sensor to the token space of an \ac{LLM}.
In either case, the underlying design remains the same as when adding listening capabilities.

The use of multimodal information can bring extended benefits to audio models.
During pretraining, it allows to substantially scale the amount of available data: while it may be hard to find sources with coupled audio and text material, video resources with rich audio-visual correspondence abound, and these can be exploited to vastly expand the quantity and diversity of pretraining data, especially for natively multimodal models.

Moreover, multimodality can be beneficial during the evaluation phase as well.
Data from vision, accelerometer, or other sensors can vastly assist in the understanding of audio.
This has already been shown by traditional methods -- e.\,g., audio-visual scene classification yielded substantially better result than only using audio to classify the soundscape~\citep{Wang21-ACD}.

\subsection{\add{Limitations of foundation models}}
\add{
We note that audio {\acp{FM}} should not be seen as a \emph{panacea} for all problems plaguing the field of audio analysis.
Instead, they constitute just another powerful tool in the array of tools available to model developers.
As such, they come with their own limitations.
A crucial one is their dependence on data and compute, an issue we return to in \mbox{\cref{sec:efficient}}.
A direct corollary of this problem is the consolidation of {\acp{FM}} in a handful of industrial organisations with the resources to train them~\mbox{\citep{Bommasani21-OTO}}.
Moreover, given their larger size (for both \emph{Tier-1} and \emph{Tier-2} {\acp{FM}}) and multimodality (for \emph{Tier-2} {\acp{FM}}), they further exacerbate the \emph{opaqueness} of {\acp{DNN}}.
{\acp{FM}} are often more complex than their traditional counterparts and thus harder to interpret.
As a result, {\acp{FM}} take away a lot of the control from the hands of model creators, which necessarily leads to a loss of accountability and trustworthiness.
}

\section{Next frontiers}
\label{sec:frontiers}
After finishing the overview of the basic principles that audio \acp{FM} are based on, we outline some of the most exciting next frontiers that arise after the successful introduction of \acp{FM} in the field of computer audition.

\subsection{Large-scale benchmarking}
The relatively recent appearance of \aclp{FM}~\citep{Bommasani21-OTO}, and the rapid experimentation that the field is currently experiencing, mean that the community has not yet settled on a common benchmark for audio \acp{FM}.
While benchmarks \add{such as \emph{HEAR} and \emph{SUPERB}} do exist for comparing audio representation learning~\citep{Turian22-HHE} and speech understanding~\citep{Yang21-SSP}, no such benchmark exists for general audio analysis \add{with \emph{Tier-2} {\acp{FM}}}.
However, as the comparison in \cref{ssec:existing} shows, it is imperative to establish such a benchmark sooner, rather than later, as there is a pressing need to understand the role of pretraining data, tasks, architectures, and training regiments in order to make consistent progress towards better, and more reliable, audio \acp{FM}.
Moreover, a common evaluation suite will facilitate the comparability of different models, like the \emph{Open-LLM-Leaderboard}\footnote{\url{https://huggingface.co/spaces/open-llm-leaderboard/open_llm_leaderboard}} does for \acp{LLM}.
Such benchmarks will be proposed soon, as their lack is readily apparent.

The designation of proper design criteria is beyond the scope of our present work.
Nevertheless, we highlight the need for proper use of datasets, given our discussion in \cref{tab:datasets}.
Specifically, any benchmark will need to avoid the use of datasets who share a common lineage in both the training and the evaluation set; for instance, derivatives of AudioSet should be used to evaluate models trained on the original AudioSet, as this cannot preclude data leakage.
Moreover, issues such as the amount of polyphony should be taken into account; while multiple datasets feature a low number of sources, real-life soundscapes can exhibit high amounts of polyphony; it is thus necessary to both train and evaluate in such conditions.
Ultimately, we expect that benchmarks will appear soon, but will be eventually superseded by newer versions -- as happens in numerous other fields.

\add{
Finally, the \emph{robustness} of models to perturbations is another interesting facet of future evaluation protocols.
\emph{Tier-1} models have already shown to be more robust to noise, at least in the case of speech~\mbox{\citep{Wagner23-DOT}}.
However, it is not clear how to conceptualise perturbations in the context of \emph{Tier-2} which are trained to holistically describe audio snippets.
There is a connection to \emph{audio retrieval}~\mbox{\citep{Xie22-LAR}}, which aims to identify the parts of an audio that are related to a (text) query.
}

\subsection{Audio generation}
\label{ssec:generation}

\Aclp{FM} are \emph{generative} models.
This naturally raises the question of whether and how these models can be used for generating audio.
While a deep dive into this topic is beyond the scope of our present work, we briefly touch on it in this subsection.

As mentioned in \cref{ssec:preliminaries}, the output tokens of a \ac{FM} must not necessarily be interpreted as a linguistic response.
In our case, they can be interpreted as audio (or, to simplify the task for the \ac{FM}, be used as input to a cascade decoder that maps them to audio).
In fact, this is how natively multimodal \acp{FM} are trained; in order to perform \ac{SSL} for the audio modality, their tokens must be interpretable as audio tokens as well.
Therefore, the extension of `fully' multimodal \acp{FM} which can both process \emph{and} output audio is a straightforward adaptation of the principles we have already discussed.
\add{This is already being exploited by {\acp{FM}} which resynthesise audio tokens to generate output audio~\mbox{\citep{Yang23-UAA, Vyas23-AUA}}.}
\add{Creating models that are capable of generating audio further allows for utilising this synthesised data in training a next iteration of {\acp{FM}}~\mbox{\citep{Meng22-GTD}}.
However, this process should be used with care as recursively training {\acp{FM}} on synthesised data can result in `model collapse'~\mbox{\citep{Shumailov24-AMC}}.}

Perhaps the most intriguing application of this capability to computer audition systems, in particular, is its use for providing \emph{sonified explanations}~\citep{Schuller21-TSI, Paissan24-LMF}.
Explanations can be useful in improving the interpretability and trustworthiness of \ac{AI} systems, especially given their propensity to err in different ways than humans.
While \acp{FM} with text components are able to provide textual explanations, extending them with audio explanations might be critical for audio models.
This is a relatively under-researched field of computer audition in general, so we expect \acp{FM} to be on the forefront of innovation given their ability to seamlessly combine audio and text responses.

\subsection{Multimodal merging}
\label{ssec:merging:multimodal}

We discussed the merging of unimodal \acp{FM} in \cref{ssec:appropriate}.
While the technique can substantially improve the capabilities of the resulting \ac{FM} (which leverages the abilities of multiple `parent' \acp{FM}), it still results in a unimodal \acp{FM}.
Therefore, to achieve multimodality, one must also connect the different components in the ways described in \cref{sec:afm}.
Here, we consider efforts and ideas on how \emph{\acp{FM} from different modalities} can be merged, thus unlocking multimodal capabilities \emph{even without additional training}. Given multimodal models $\mathcal{M}_1$ and $\mathcal{M}2$ of the same type, the challenge is to produce a new model $\mathcal{M}_{3}$ that incorporates the abilities of both preceding models.

\textbf{Merging vision-language models:}
A significant challenge in merging \acp{VLM} is ensuring that the model can seamlessly switch between and integrate the different modalities. This requires sophisticated training regimes that balance the contributions of each modality and ensure that neither dominates the other. Successful merging of \acp{VLM} results in models capable of tasks such as image captioning, visual question answering, and generating images from textual descriptions. Existing methods such as \ac{EMA}~\citep{Cai21-EXP}, ``Model Soups'', and Fisher-weighted averaging have shown effectiveness but struggle with the complexities of \acp{VLM}~\cite{Matena22-MER}.
To address these challenges, \citet{Ye23-MER} proposed a gating network. This method can merge all layers within a \ac{VLM} (e.\,g., embedding, normalisation, attention, and \ac{MLP}) and select the appropriate classifier. Trained on unlabelled datasets from all tasks, the gating network predicts which task the input belongs to and merges the models during inference. Additionally, the authors designed a novel metric of model weight similarity to boost performance, especially because merging tasks increases in difficulty. Another recent study presented the \ac{EMR}-merging~\citep{Huang24-EMR} method while showing superior performance over existing merging methods in both traditional and new scenarios. This encompasses the merging of numerous vision models (up to 30), \ac{NLP} models, \ac{PEFT} models, and multi-modal models.

\textbf{Merging audio-language models:}
The concept of merging \acp{ALM} is expected to bring a major advancement to audio analysis. Like \acp{VLM}, \acp{ALM} will integrate multiple types of data, in this case, audio and text. Techniques from \acp{VLM}, such as model averaging, can be adapted to merge language and audio models, allowing these to handle diverse audio data effectively. By leveraging pretrained models, and aligning them with methods similar to those used in vision-language merging, the need for extensive retraining is reduced. This will result to \acp{ALM} capable of interpreting complex audio-language content. Although the \remove{domain of }merging \add{of} \acp{VLM} and \acp{ALM} \change{is not as widely researched as that of {\acp{LLM}}}{has not been attempted yet}, it is anticipated \change{to become increasingly sophisticated with new methods and innovations capable of performing complex tasks}{to be one of the next steps for improving audio {\acp{FM}}}. 

\subsection{Beyond human evaluations}
A major proposition of \acp{FM} is their ability to exhibit \emph{emergent abilities} as a byproduct of a) their scale and b) the quantity of data they have been trained on.
We discussed some limitations of the datasets where existing models are being trained on in \cref{ssec:datasets}.
For the purposes of this section, we assume that future \acp{FM} will be trained on vastly larger resources, some of which contain audio that goes beyond human capabilities of listening.
For instance, huge (largely unlabelled) corpora exist from ecoacoustic monitoring projects, where large environmental areas are passively monitored using audio sensors~\citep{Farina17-ECO, Schuller23-ECA}.
Oftentimes, these sensors record audio in ultrasonic ranges~\citep{Walters13-CHA}.

This raises the question of how \acp{FM} will incorporate aspects of these audio sources in their capabilities.
Moreover, it introduces the challenge of measuring these capabilities.
Human hearing is not the only type of hearing in nature -- in fact, depending on the `target', it is not even the best one.
For example, some animals have much \change{longer}{larger} hearing ranges than humans, and are better able to distinguish between tones and rhythms.

\acp{FM} may, presumably, exhibit such capabilities as well.
However, we are entirely lacking ways to evaluate such competences.
Auditory benchmarks (and most \ac{AI} benchmarks for that matter) are almost exclusively anthropocentric -- their `ground truth' labels are derived from human annotations and essentially encode human perception.
This makes them unsuited to judge performance that exceeds that of humans.

Extending benchmarks to incorporate alternative notions of performance -- perhaps motivated by animals -- is not an easy feat.
We expect more research to be focused on this direction, e.\,g., by monitoring animals in their native environments and learning to associate their responses with hearing skills, or incorporating external sensors as an extra verification step for machine hearing abilities that outperform those of humans.
Ultimately, the issue of superhuman performance for \acp{FM} geared towards audio analysis is not as critical as that in other domains~\citep{Bommasani21-OTO, Triantafyllopoulos24-EAS}, but it nevertheless represents an exciting new frontier for audio researchers.

\subsection{\add{(In)Efficient foundation models}}
\label{sec:efficient}
\add{
A fundamental limitation of contemporary {\acp{FM}} is their inefficiency with regards to both data and compute resources.
This raises the `barrier-of-entry' for researchers without access to the compute resources and datasets to train large models.
For instance, the \emph{Llama-3-405B} model was trained on ``up to 16K Nvidia H100 GPUs'' and requires 16 GPUs for inference~\mbox{\citep{Grattafiori24-TL3}}.
This is far and beyond what most researchers have available.
It is also far beyond the requirements for many real-world applications which aim to deploy audio analysis models on edge devices~\mbox{\citep{Schmid24-DLA}} and imparts a high environmental cost~\mbox{\citep{Wu22-SAE}}.
\emph{Pruning}~\mbox{\citep{Blalock20-WIT}}, \emph{knowledge distillation}~\mbox{\citep{Gou21-KDA}}, or \emph{early-exit models}~\mbox{\citep{Rahmath24-EDN}} are all possible pathways to achieve a reduction in computational complexity during inference.
For \emph{Tier-2} audio {\acp{FM}}, in particular, there is substantial promise in reducing the computational overhead of the \emph{text decoder}, as these models need, presumably, a much less powerful decoder than general-purpose {\acp{LLM}}.
}

\subsection{\add{Doing research in the era of foundation models}}
\add{
We end with a discussion on the future of research in audio analysis.
The trends we have outlined so far seem to coalesce into a future where {\acp{FM}} dominate the field due to their higher predictive performance and ease-of-use, at least in application domains where the computational budget can accommodate them.
However, this does not preclude the further development of traditional pipelines and their consolidation into more complex {\acp{FM}}.
This is a trend also observed in the application of deep learning to audio analysis.
Oftentimes, components from traditional signal processing were `revived' and integrated into {\acp{DNN}}~\mbox{\citep{Engel20-DDD, Zeghidour21-LAL}}.
A similar fate may await existing pipelines which now seem `under threat' from the adoption of {\acp{FM}}.
We thus caution against a complete abandonment of traditional models, at least as it pertains to the education of newcomers to the field.
}

\section{Conclusion}
\label{sec:conclusion}
We have presented an overview of how the field of computational audio analysis is rapidly transitioning from traditional, monolithic, and task-specific pipelines to modular, multi-tasking \aclp{FM}.
This is largely happening by connecting audio modules to existing \aclp{LLM} which have been pretrained on large text corpora and thus encapsulate a large amount of world knowledge -- although `natively' multimodal \acp{FM} are also beginning to appear.
These models can handle a combination of audio and language (and even more) inputs, and leverage their extensive pretraining to provide unprecedented generalisation capabilities.
Moreover, given the ingrained use of language, they offer a more convenient and intuitive interface for human users through the use of text queries.
Finally, they are paving the way towards a consolidation of computer audition tasks.
Rather than using independent pipelines to achieve the goals of each task, the community is transitioning to a paradigm where these capabilities are \emph{unlocked} in audio \acp{FM} through the use of (instruction) finetuning.
Crucially, unlike `traditional' transfer learning, these newly-unlocked capabilities can co-exist with ones needed for other tasks.
This new paradigm can be thus concisely described as the quest to find the `one model to rule them all'.

Our overview presents the basic principles behind traditional pipelines and how those are upended by \acp{FM}.
We have outlined the key components needed to add `hearing' capabilities to pretrained \acp{LLM} and how these capabilites can be augmented using further finetuning or \acl{RAG}.
Our emphasis on key principles aims to inform the audio community on these recent, exciting advances, highlight the peculiarities of audio analysis for the broader \ac{AI} community working on \acp{FM}, and motivate the development of the next wave of audio \acp{FM} and their incorporation to the `daily life' of audio researchers.

\section*{Acknowledgments}
This work was partially funded from the DFG's Reinhart Koselleck project No.\ 442218748 (AUDI0NOMOUS).

\section{\refname}
 \printbibliography[heading=none]

\appendix

\section{Existing audio foundation models (Full Table)}
\label{app:existing}

\cref{tab:overview-full} shows the complete list of datasets and audio-related abilities for the audio-language models included in our work, including speech and music.
However, we still exist other modalities or general linguistic capabilities.

\begin{landscape}
\begin{table}[t]
\centering
\caption{Overview of existing \emph{Tier-2} audio-language models including tasks beyond computer audition.}
\label{tab:overview-full}
\begin{threeparttable}
\resizebox{\textwidth}{!}{\begin{tabular}{p{0.2\textwidth}|p{0.3\textwidth}p{0.4\textwidth}p{0.1\textwidth}|p{0.4\textwidth}}
\toprule
\textbf{Model} & \textbf{Modules} & \textbf{Data Used} & \textbf{Fusion Type} & \textbf{Capabilities}\\
\midrule
\multicolumn{5}{c}{\textbf{Publicly Available}}\\
\midrule
Pengi~\citep{Deshmukh23-PAA} & Audio encoder\textcolor{orange}{\faFire}: CLAP, \newline Text encoder\textcolor{blue}{\ding{100}}: CLAP \newline Audio and Text Prefix Layers\textcolor{orange}{\faFire} \newline LLM\textcolor{blue}{\ding{100}}: GPT2-base & AudioSet, FSD50K, CochlScene, MSP-Podcast, CMU-MOSI, CMU-MOSEI, MELD, NSynth, FMA, AudioCaps, Clotho (v2), ClothoAQA, WavText5K, SoundDescs, MACS, WavCaps, Freesound, FindSound & Early & Audio captioning, Audio question answering, Sound event classification, Music analysis, Instrument classification, Music note analysis, Acoustic scene classification, Speech emotion recognition, Vocal sound classification, Audio-language retrieval\\
\midrule
APT-LLM~\citep{Liang23-APT} & Audio encoder: Audio-MAE\textcolor{blue}{\ding{100}} \newline Audio aligner\textcolor{orange}{\faFire} \newline 
Acoustic adapters\textcolor{orange}{\faFire}: APT \newline LLM\textcolor{blue}{\ding{100}}: Vicuna-7b (v1.1)$^\dag$ & AudioSet, Wavcaps, AudioCaps, Clotho (v2), Clotho AQA, AudioSet-SL, NLAR & Early & Audio tagging, Sound event classification, Natural language-audio reasoning, Audio-visual question answering\\
\midrule
LTU~\citep{Gong24-LTA} & Audio encoder\textcolor{orange}{\faFire}: AST \newline LoRA adapters\textcolor{orange}{\faFire} \newline LLM\textcolor{blue}{\ding{100}}: LLaMA-7b & OpenAQA & Early & Audio tagging, Audio captioning, Audio reasoning, Audio temporal analysis, Sound event classification
\\
\midrule
LTU-AS~\citep{Gong23-JAA} & Audio encoder\textcolor{blue}{\ding{100}}: Whisper \newline TLTR\textcolor{orange}{\faFire} \newline
Projection layer\textcolor{orange}{\faFire}\newline LoRA adapters\textcolor{orange}{\faFire} \newline LLM\textcolor{blue}{\ding{100}}: LLaMA-7b & Open-ASQA & Early & Audio captioning, Sound event classification, Audio question answering, Gender classification, Age prediction, Speech recognition, Sound event classification\\
\midrule
QWEN-Audio~\citep{Chu23-QAU} & Audio encoder\textcolor{orange}{\faFire}: Whisper-L-v2, \newline LLM\textcolor{blue}{\ding{100}}: Qwen-7b & CoVOST 2, Macaw-LLM Instruction Dataset, MosIT, LLaSM-Audio-Instructions Dataset 
& Early & Audio captioning, Sound event classification, Acoustic scene classification, Sound event detection, Audio question answering \\
\midrule
Salmonn~\citep{Tang24-STG} & Audio encoders\textcolor{blue}{\ding{100}}: Whisper, BEATs, 
\newline Q-Former\textcolor{orange}{\faFire} \newline 
LoRA adapters\textcolor{orange}{\faFire} \newline
LLM\textcolor{blue}{\ding{100}}: Vicuna & LibriSpeech, GigaSpeech M-set, WavCaps, AudioCaps, MusicCaps, Clotho, VoxCeleb1, MillionSOng, MusicNet & Early & Audio captioning, Speech recognition, Speech emotion recognition, Speaker verification, Speech question answering, Audio question answering, Music question answering, Phone recognition, Gender recognition\\
\midrule
BAT~\citep{Zheng24-BLT} & Audio encoder\textcolor{blue}{\ding{100}}: Spatial-ASL, \newline Projection Layer\textcolor{orange}{\faFire} \newline LLM\textcolor{orange}{\faFire}: LLaMA2 & SpatialSoundQA &  Early & Sound event classification, Sound event detection and spatial localisation, Audio question answering\\
\midrule
Audio Flamingo~\citep{Kong24-AFA} & Audio encoder\textcolor{blue}{\ding{100}}: CLAP \newline Projection Layer\textcolor{orange}{\faFire} \newline LLM\textcolor{orange}{\faFire}: opt-iml-max-1.3b & WavCaps, Macs, SoundDescs, Clotho-v2, WavText5K, LAION-630k, Clotho-AQA, Open-AQA, AudioSet, FSD50k, CochlScene, NonSpeech7K, Chime-Home, Sonyc-UST, LP-MusicCaps, MusicCaps, MusicQA, MusicAVQA, NSynth, MTG-Jamendo, FMA, MusDB-HQ, MSP-Podcast, Emov-DB, JL-Corpus, Tess, MELD, OMGEmotion & Deep & Audio captioning, Sound event classification, Audio question answering \\
\midrule
UniAudio 1.5~\citep{Yang23-UAA} & Audio encoder\textcolor{orange}{\faFire}: LLM-Codec \newline LLM\textcolor{blue}{\ding{100}}: LLaMA2-7b & MLS, AudioCaps & Early & Sound event classification, Speech emotion classification, Text-to-speech generation, Speech enhancement\\
\midrule

\multicolumn{5}{c}{\textbf{Restricted Access}}\\
\midrule
Gemini 1.5 Pro~\citep{Reid24-GUM} & N/A & N/A & N/A & Audio understanding, Audio captioning, Spoken language understanding$^*$\\
\midrule
GPT-4o\footnote{\url{https://openai.com/index/hello-gpt-4o/}} & N/A & N/A & N/A & Audio understanding, Audio captioning, Spoken language understanding$^*$\\
\midrule
LauraGPT~\citep{Chen23-LLA} & Audio encoder\textcolor{orange}{\faFire}: Encodec-based \newline LLM\textcolor{orange}{\faFire}: Qwen & AISHELL-1, AISHELL-2, WenetSpeech, LibriSpeech, GigaSpeech, SLURP, BSTC, CoVOST 2, MELD, IEMOCAP, RAVDESS, TESS, Crema-D, Emov-DB, SAVEE, Clotho, AudioCaps, WavCaps, WSJ, FSD-50K, RIR, LibriTTS, 3D-Speaker, ParaCrawl & Early & Automatic speech recognition, Spoken language understanding, Speech to text translation, Speech emotion recognition, Automated audio captioning, Speech enhancement, Text-to-speech synthesis, Machine translation\\
\bottomrule
\end{tabular}}
\begin{tablenotes}
    \item[*] Only showing audio-related tasks as these models support multiple modalities and skills.
\end{tablenotes}
\end{threeparttable}
\end{table}
\end{landscape}


\begin{IEEEbiography}[{\includegraphics[width=1in,height=1.25in,clip,keepaspectratio]{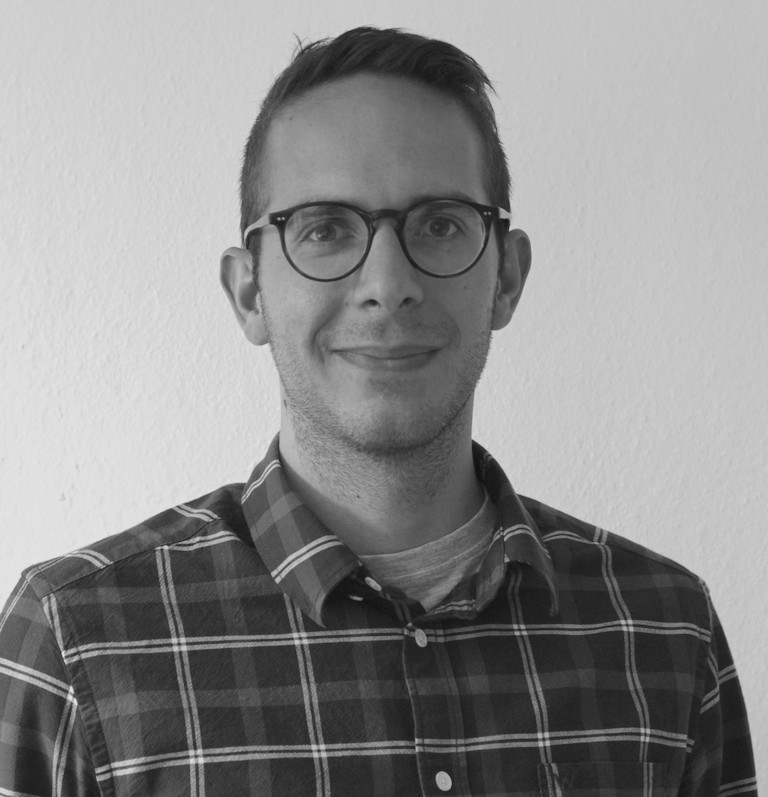}}]{Andreas Triantafyllopoulos}
is a research assistant and doctoral candidate at the Technical University of Munich. He received his M.Sc. in Electrical and Computer Engineering from the University of Patras in 2017. He has been working on deep learning for audio understanding, with an emphasis on speech analysis, since 2018, when he joined audEERING GmbH as an AI researcher. He previously worked as a research assistant at the University of Augsburg.
\end{IEEEbiography}

\begin{IEEEbiography}[{\includegraphics[width=1in,height=1.25in,clip,keepaspectratio]{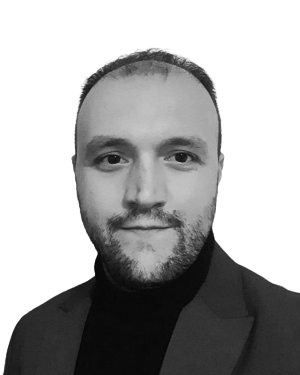}}]{Iosif Tsangko} received the B.Sc. and M.Sc. degrees in Mathematics from the University of Crete, Heraklion, Greece, in 2016 and 2018, respectively, and an M.Sc. in Computer Science from the Hellenic Army Academy, Athens in 2022. He is currently pursuing a Ph.D. at the Technical University of Munich, Munich, Germany, with a focus on machine learning, AI, and health informatics. His research interests include speech enhancement, large language models, and AI applications in health.
\end{IEEEbiography}

\begin{IEEEbiography}[{\includegraphics[width=1in,height=1.25in,clip,keepaspectratio]{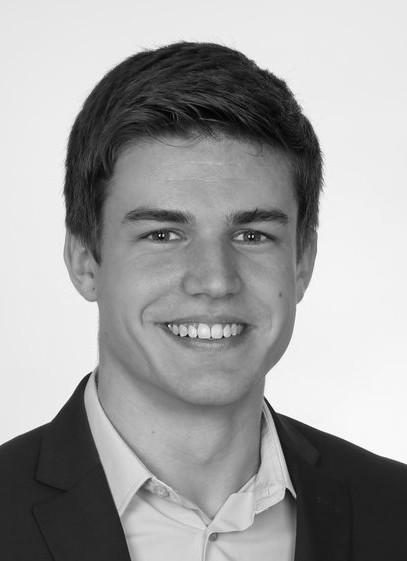}}]{Alexander Gebhard}
is a research assistant and doctoral candidate at the Technical University of Munich. He received his M.Sc. in Computer Science from the Friedrich-Alexander University of Erlangen-Nuremberg (FAU) in 2020. He has been working on deep learning for audio understanding since 2021, with an emphasis on ecoacoustics. He previously worked as a research assistant at the University of Augsburg.
\end{IEEEbiography}

\begin{IEEEbiography}[{\includegraphics[width=1in,height=1.25in,clip,keepaspectratio]{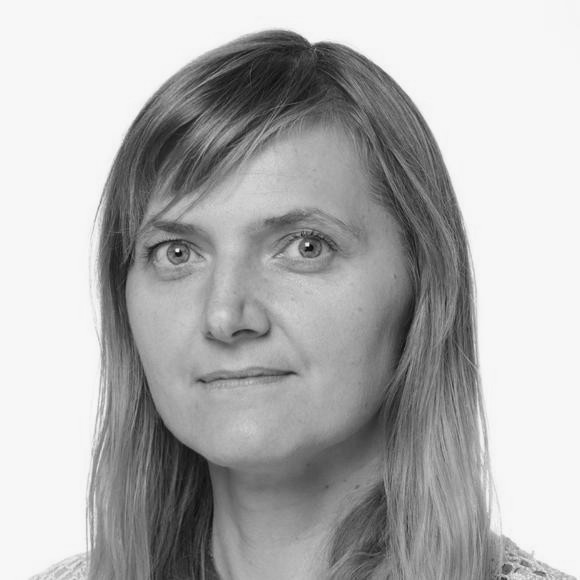}}]{Annamaria Mesaros}
is an Associate Professor at Tampere University. She received her PhD in Signal Processing at Tampere University of Technology in 2012. Her research focuses on sound event detection in real-world multisource environments and includes over 50 scientific publications and many open datasets. She is an IEEE Senior Member and member of the Audio and Acoustic Signal Processing Technical Committee of IEEE Signal Processing Society. She is a coordinator of the Detection and Classification of Acoustic Scenes and Events (DCASE) Challenge, vice-chair of the DCASE Steering Group, and currently an Academy of Finland Research Fellow for ``Teaching Machines to Listen''.
\end{IEEEbiography}

\begin{IEEEbiography}[{\includegraphics[width=1in,height=1.25in,clip,keepaspectratio]{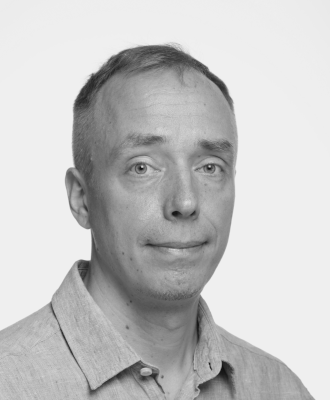}}]{Tuomas Virtanen}
 is Professor at Tampere University, Finland, where
he is leading the Audio Research Group. He received the M.Sc. and
Doctor of Science degrees in information technology from Tampere
University of Technology in 2001 and 2006, respectively. He has
also been working as a research associate at Cambridge University
Engineering Department, UK. He is known for his pioneering work on
computational acoustic scene analysis and sound source separation.
His research interests include machine listening, computational content
analysis of audio, and machine learning for audio. He has authored more
than 200 scientific publications on the above topics, which have been
cited more than 20000 times. He has received IEEE Signal Processing
Society best paper awards multiple times, as well as many other best
paper awards. He is an IEEE Fellow, IEEE Signal Processing Society
Distinguished Lecturer 2024-2025, recipient of the ERC 2014 Starting
Grant, and has been a member of the Audio and Acoustic Signal Processing
Technical Committee of IEEE Signal Processing Society.
\end{IEEEbiography}

\begin{IEEEbiography}[{\includegraphics[width=1in,height=1.25in,clip,keepaspectratio]{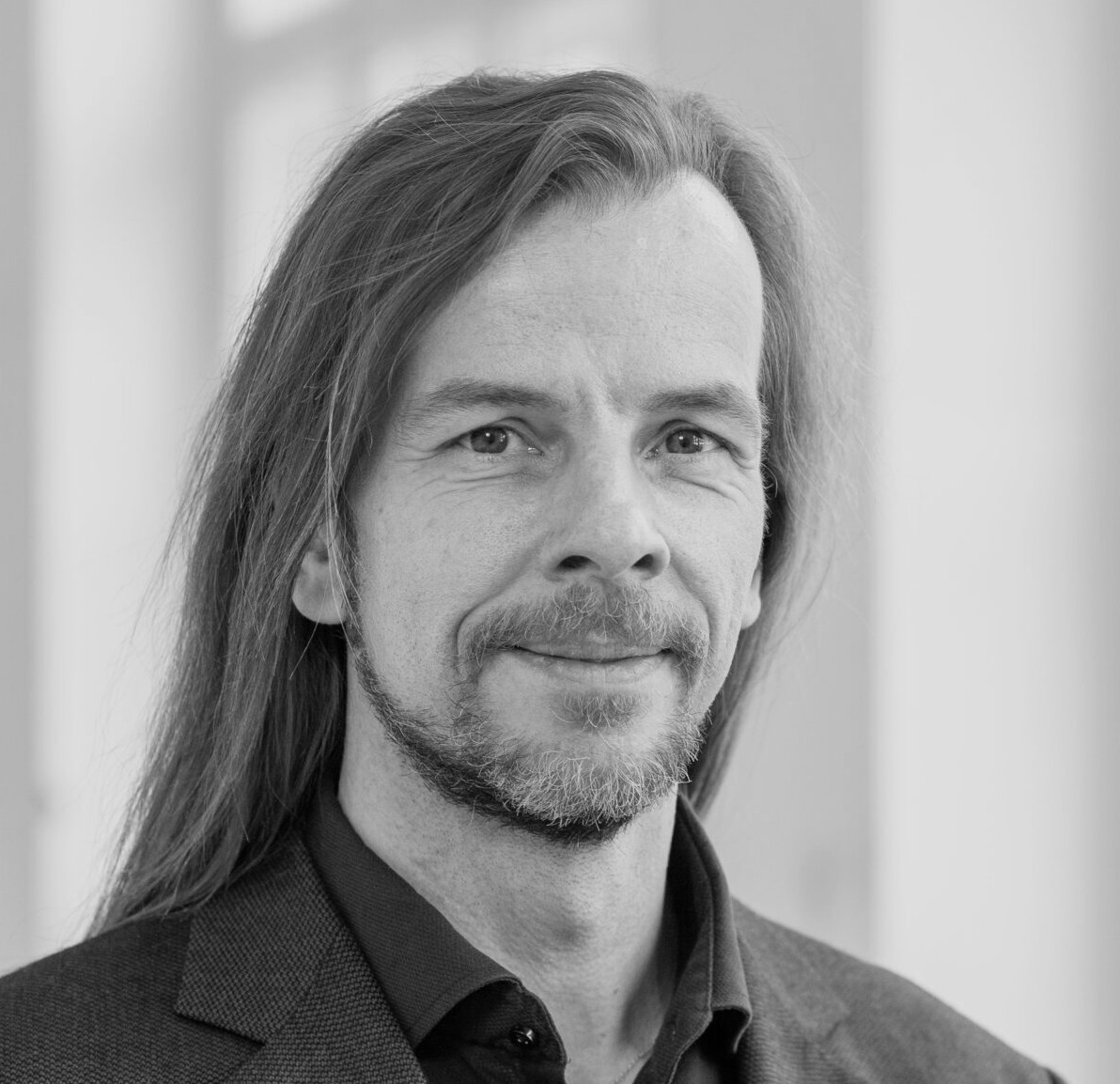}}]{Björn Schuller}
is Full Professor and Chair of Health Informatics at the University Hospital of the Technical University of Munich. He is also Full Professor of Artificial Intelligence and the Head of GLAM at Imperial College London/UK, co-founding CEO and current CSO of audEERING -- an Audio Intelligence company based near Munich and in Berlin/Germany, Core Member in the Munich Data Science Institute (MDSI), Principal Investigator in the Munich Center for Machine Learning (MCML), and Fellow of the Imperial Data Science Institute, amongst other Professorships and Affiliations. He is a Fellow of the ACM, Fellow of the IEEE and Golden Core Awardee of the IEEE Computer Society, Fellow of the BCS, Fellow of the ELLIS, Fellow of the ISCA, Fellow and President-Emeritus of the AAAC, and Elected Full Member Sigma Xi. He (co-)authored 1,500+ publications (75,000+ citations, h-index 121 ranking him number 8 in the UK for Computer Science), is Field Chief Editor of Frontiers in Digital Health, Editor in Chief of AI Open and was Editor in Chief of the IEEE Transactions on Affective Computing amongst manifold further commitments and service to the community. His 50+ awards include having been honoured as one of 40 extraordinary scientists under the age of 40 by the WEF in 2015. Recently, he was awarded ACM Distinguished Speaker for the term 2024-2027 and IEEE Signal Processing Society Distinguished Lecturer 2024.
\end{IEEEbiography}

\end{document}